\documentclass[prb,aps,longbibliography,floats,superscriptaddress,twocolumn]{revtex4-2}

\usepackage[utf8]{inputenc}
\usepackage{amsmath,amssymb,amsthm,amsbsy,bm}
\usepackage{graphicx,epsfig,epstopdf,subfigure}
\usepackage{dsfont,verbatim,varwidth,dcolumn}
\usepackage[usenames,dvipsnames]{xcolor}
\usepackage{color,soul,subfiles}
\usepackage[colorlinks]{hyperref}
\usepackage[normalem]{ulem}
\usepackage{BLGNu2}

\newcommand\Ds{D^{*}}
\newcommand\Bsa{B_{s}}
\newcommand\Bsb{B_{*}}

\makeatother

\begin{document}

\title{Phase diagram of the $\nu = 2$ quantum Hall state in bilayer graphene}
\author{Udit Khanna}
\affiliation{Department of Physics, Bar-Ilan University, Ramat Gan 52900, Israel}
\author{Ke Huang}
\affiliation{Department of Physics, The Pennsylvania State University, University Park, Pennsylvania 16802, USA}
\author{Ganpathy Murthy}
\affiliation{Department of Physics and Astronomy, University of Kentucky, Lexington, Kentucky 40506, USA}
\author{H.~A. Fertig}
\affiliation{Department of Physics, Indiana University, Bloomington, Indiana 47405, USA}
\author{Kenji Watanabe} 
\affiliation{Research Center for Functional Materials, National Institute for Materials Science, 1-1 Namiki, Tsukuba 305-0044, Japan}
\author{Takashi Taniguchi}
\affiliation{International Center for Materials Nanoarchitectonics, National Institute for Materials Science, 1-1 Namiki, Tsukuba 305-0044, Japan}
\author{Jun Zhu}
\affiliation{Department of Physics, The Pennsylvania State University, University Park, Pennsylvania 16802, USA}
\author{Efrat Shimshoni}
\affiliation{Department of Physics, Bar-Ilan University, Ramat Gan 52900, Israel}

\begin{abstract}
Bilayer graphene exhibits a rich phase diagram in the quantum Hall regime, arising from a multitude of internal degrees of freedom, including spin, valley, and orbital indices. The variety of fractional quantum Hall states between filling factors $1 < \nu \leq 2$ suggests, among other things, a quantum phase transition between valley-unpolarized and polarized states at a perpendicular electric field $\Ds$. We find the behavior of $\Ds$ with $\nu$ changes markedly as $B$ is reduced. At $\nu = 2$, $\Ds$ may even vanish when $B$ is sufficiently small. We present a theoretical model for lattice-scale interactions which explains these observations; surprisingly,  both repulsive and attractive components in the interactions are required. Within this model we analyze the nature of the $\nu = 2$ state as a function of the magnetic and electric fields, and predict that valley-coherence may emerge for $D \sim \Ds$ in the high $B$ regime. This suggests the system supports Kekule bond-ordering, which could in principle be verified via STM measurements.
\end{abstract}

\maketitle

\textit{Introduction.}~The quantum Hall (QH) regime of two-dimensional electronic systems with several internal degrees 
of freedom presents an intriguing many-body problem, where the interplay of interactions and degenerate 
Landau levels (LLs) often leads to a multitude of possible ground 
states~\cite{PrangeGirvin,GirvinMacDonald,Ezawa,HalperinJain,ValleyNematic_Review}. 
Graphene and its few-layer variants offer compelling material platforms to explore this interplay due to their rich Landau 
spectrum, involving approximate SU(4) symmetry in spin and valley sectors, as well as relatively high mobilities and wide gate 
tunability~\cite{CastroNeto_Graphene_RMP,KunYang_PRB_2006,PKim_PRL_2006,PKim_PRL_2007,3LG_2013,3LG_2017,4LG_2020}. 

Graphene systems, uniquely, support QH phases around charge neutrality, whose nature has been investigated extensively. 
Previous theoretical studies~\cite{AliceaFisher_2006,Kharitonov_MLG_2012,Kharitonov_BLG_2012,Kharitonov_BLG_PRB_2012,Braden_MLG_2016,GM_BLG_2017}  
have clarified that the order underlying the ground state depends crucially on lattice-scale 
corrections to the (long-range) Coulomb interaction, which reduce the valley SU(2) symmetry to U(1)$\times \mathbb{Z}_2$.
The precise form of these corrections is unclear and may depend on the device configuration.  
In light of this, the standard approach, introduced by Kharitonov~\cite{Kharitonov_MLG_2012,Kharitonov_BLG_2012,Kharitonov_BLG_PRB_2012}, 
is to include phenomenological terms consistent with the symmetry. Conventionally, these terms are assumed to be independent of the magnetic field ($B$) and 
have a range of order the lattice constant, 
which is much smaller than the magnetic length [$\ell = \sqrt{\hbar / e B}$]. 
In what follows we will refer to this as the orthodox model (OM) of the lattice scale interactions. 

Generally, the OM has been in accordance with  experimental observations. 
In particular, for the $\nu = 0$ phase of monolayer and bilayer graphene (MLG and BLG respectively), this model  supports the interpretation of transport~\cite{PKim_BLG_2010,PKim_MLG_2012,PKim_BLG_Nu0_2013,Young_MLG_2014,Jun_BLG_Nu0_2019}  
and magnon transmission~\cite{AYacoby_MLG_2018,MLG_CAF_2018,Jun_CAF_2021} experiments in terms of a magnetically ordered 
ground state. However, recent scanning tunneling measurements~\cite{STM_1,STM_2,STM_3} in MLG find charge-ordered ground states at $\nu = 0$, 
with a Kekule bond-order (BO) or a charge density wave (CDW) order. 
Given the difficulty in reconciling these conflicting observations (within the OM), recently one of 
us~\cite{GM_PRL_2022,GM_Nu0_2022} reevaluated the $\nu = 0$ phase diagram allowing 
the lattice-scale interactions to assume a more generic form. These studies find that coexistence phases with both spin and charge order 
may appear if interactions have a structure at the scale of $\ell$. 

As a matter of principle, and irrespective of specific filling factor and device details, the 
phenomenological terms in the low energy model may have a complicated form due to 
quantum fluctuations involving other (positive and negative energy)  
LLs~\cite{GM_LLMix,Nayak_LLMix,Peterson_LLMix,Sodemann_LLMix,Simon_LLMix}. This LL mixing is largely 
controlled by the parameter  $\ka = E_{c} / \hoc$, the ratio of the Coulomb energy scale ($E_{c}$) to the cyclotron gap ($\hoc$)~\cite{Peterson_LLMix}. 
In general, LL mixing introduces a non-trivial component with a range of $\ell$ to the effective interactions, 
which may be {\it attractive or repulsive}. Moreover, the $B$-dependence of these terms may be different from that of the bare terms. 
Refs.~\onlinecite{GM_PRL_2022,GM_Nu0_2022} demonstrate that such considerations not only affect the energetics, but also add to the 
set of possible ground states~\cite{note1}.

In this Letter, we explore the QH phases of BLG and provide further evidence for the crucial role of such modified interactions. We  consider a dual-gated device, which allows the application of a transverse electric field ($D$) as an experimental knob to tune between different ground states at fixed filling factor $\nu$. 
Close to charge neutrality, the chemical potential lies within a set of eight (nearly) degenerate LLs 
labelled by spin, valley and an orbital index, supporting a variety of broken-symmetry states in the range $|\nu| < 4$. 
Indeed, transport~\cite{PKim_FQH_BLG_2014,Jun_BLG_LLs,Jun_BLG_Nu0_2019,Jun_BLG_PRX2022} and  
capacitance~\cite{Hunt_BLG_2017} measurements provide evidence for a complex sequence of phase transitions driven by $D$ 
for both integer and fractional fillings. The number of transitions and the values of $D$ at which they occur are functions of $\nu$ and $B$. As shown below, the complete phase diagram in the \{$D$, $B$, $\nu$\} space encodes vital information on the underlying many-body effects.

Notably, the OM is consistent with many earlier measurements, restricted to certain regions of the parameter space such as a fixed value of $B$~\cite{Hunt_BLG_2017} or integer $\nu$~\cite{Jun_BLG_LLs}.
Our present study is based on transport measurements over a wide range of parameters, including moderate and high $B$, where experimental data in higher-quality samples are now available. 
Specifically, we focus on filling factors $1 < \nu \leq 2$, and track the variation of the critical electric field ($\Ds$)
at which a phase transition occurs with $B$ and $\nu$ [Fig.~\ref{fig1}]. We find that $\Ds(\nu)$ is an
increasing (decreasing) function of $\nu$ at high (low) $B$-fields, and that $\Ds$ may even vanish at sufficiently low fields. 
It is worth emphasizing that because the chemical potential is pinned to the same LL for this filling factor range, the behavior of $\Ds (\nu, B)$ is controlled by the lattice-scale interactions, and imposes significant constraints on their form. The elucidation of these interactions is the main purpose of this work.
 
The main finding of this Letter is that the OM of lattice-scale  interactions cannot account for the observed behavior of 
$\Ds(\nu, B)$. Our Hartree-Fock (HF) analysis demonstrates that 
the symmetry-breaking interactions must have both repulsive and attractive components with different $B$-dependence 
in order to explain the measurements [Fig.~\ref{fig2}]. These results suggest that corrections arising from the LL-mixing play a significant role in BLG, particularly at lower $B$. 
We further employ this model to construct the phase diagram of the $\nu = 2$ QH state 
in the $B$--$D$ plane [Fig.~\ref{fig3}]. Interestingly, we find the emergence of an inter-orbital valley-coherent phase around $D \sim \Ds$ for sufficiently large $B$. 
The existence of such a valley-orbital entangled (VOE) phase at high $B$ implies that the transport gap at $\nu = 2$ does not close around $D = \Ds$. 
Additionally, valley-coherence points to the presence of a Kekule BO phase which may be observed in tunnelling measurements, similar to those reported in Refs.~\onlinecite{STM_1,STM_2,STM_3}. 

\begin{figure}[t]
\centering
\includegraphics[width=\columnwidth]{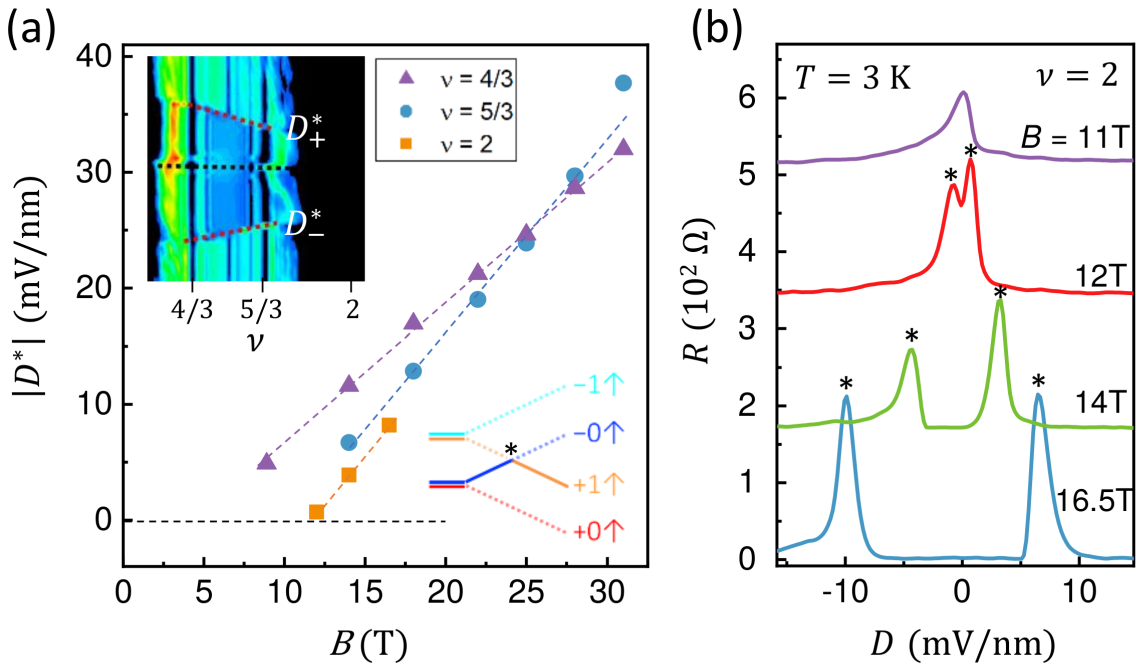}
\caption{(a) Magnetic field dependence of the critical electric field $\Ds$, at which a first-order phase transition occurs, 
for different filling factors ($\nu$) as labeled in the graph.  
The upper inset shows a false color map of $R_{xx} (\nu, D)$ between $\nu = 1$ and $2$ at $B = 18$ T 
(obtained in Ref.~\cite{Jun_BLG_PRX2022}). The red dashed lines mark the positive and negative ($\Ds_{\pm}$) 
transitions. The black dashed line marks the true $D = 0$. (b) Line scans of $R_{xx}$ vs $D$ for $\nu = 2$ at 
different $B$. The resistance peaks (marked with $*$) correspond to $\Ds_{\pm}$. 
The average of $\Ds_{\pm}$ at $\nu = 2$ is plotted as squares in (a). Using similar measurements, $\Ds_-$ for $\nu = 4/3$ ($5/3$) are obtained and shown as triangles (circles). Dashed lines are guide to the eye.}
\label{fig1}
\end{figure}

\textit{Transport Measurements.}~We employ a high quality dual-gated BLG device, device 002, described previously in Ref.~\onlinecite{Jun_BLG_PRX2022}, to examine the behavior of $\Ds$ as a function of B at different filling factors. The upper inset of Fig.~\ref{fig1}(a) shows a color map of $R_{xx} (D)$ in the range $ 1 < \nu < 2 $ at $B = 18$ T and $T = 20$ mK (See the full dataset covering a wider range of $\nu$ in Ref.~\onlinecite{Jun_BLG_PRX2022}). Regions with darker colors correspond to vanishingly small $R_{xx}$ indicating QH phases. The black dashed curve marks the true inversion symmetric line, i.e. where $D = 0$ is. Device asymmetry causes a slight asymmetry between $\Ds_{+}$ and $\Ds_{\neg}$ (the red dashed lines), where two LLs with different valley and orbital indices cross, as depicted in the lower inset ~\cite{Jun_BLG_LLs}.  Fig.~\ref{fig1}(b) plots $R_{xx} (D)$ traces taken at $T = 3$ K, where $\Ds_{+}$ and $\Ds_{\neg}$ become more readily observed for $\nu = 2$ and manifest as resistance peaks. The closing of the transport gap signals a first order phase transition, similar to previous observations in GaAs ~\cite{Shayegan_Rxx_Spike}. Their positions are marked by $*$s and evolve with B. $\Ds$ values extracted from similar measurements are plotted in the main panel of Fig.~\ref{fig1}(a) for $\nu= 2, 5/3, 4/3$.

\begin{figure*}[t]
\centering
\includegraphics[width=0.31\textwidth]{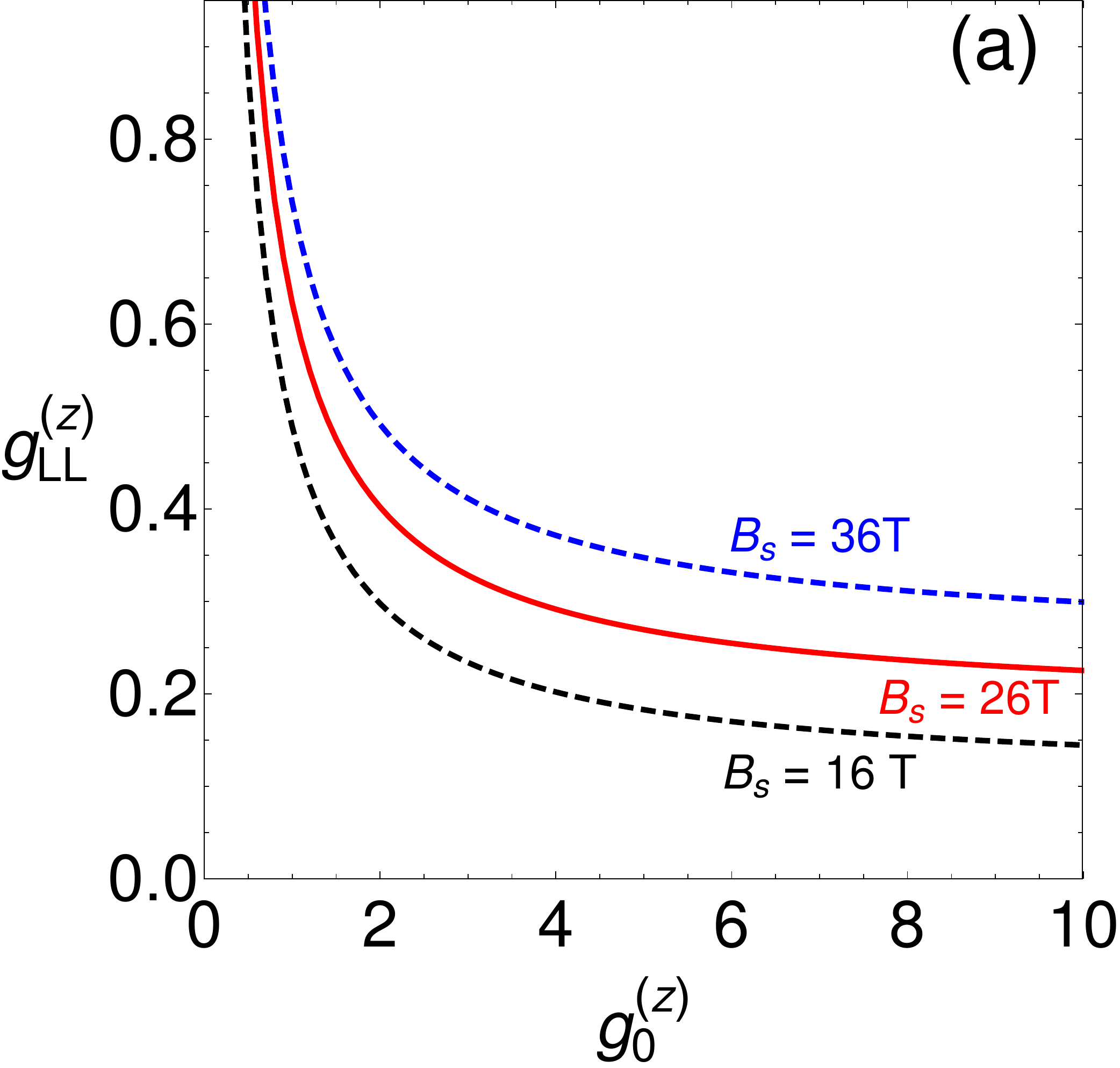}
\includegraphics[width=0.31\textwidth]{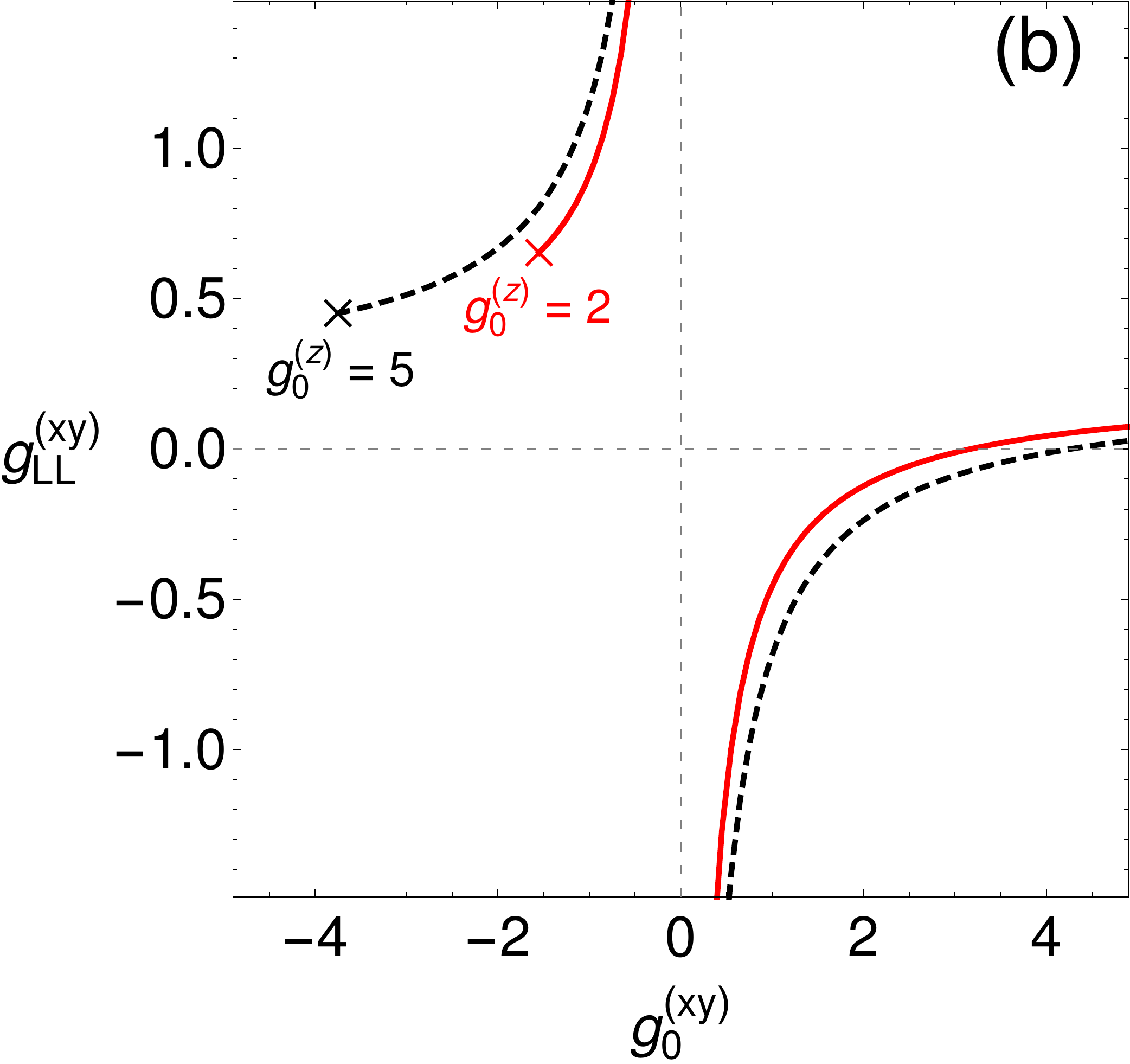}
\includegraphics[width=0.36\textwidth]{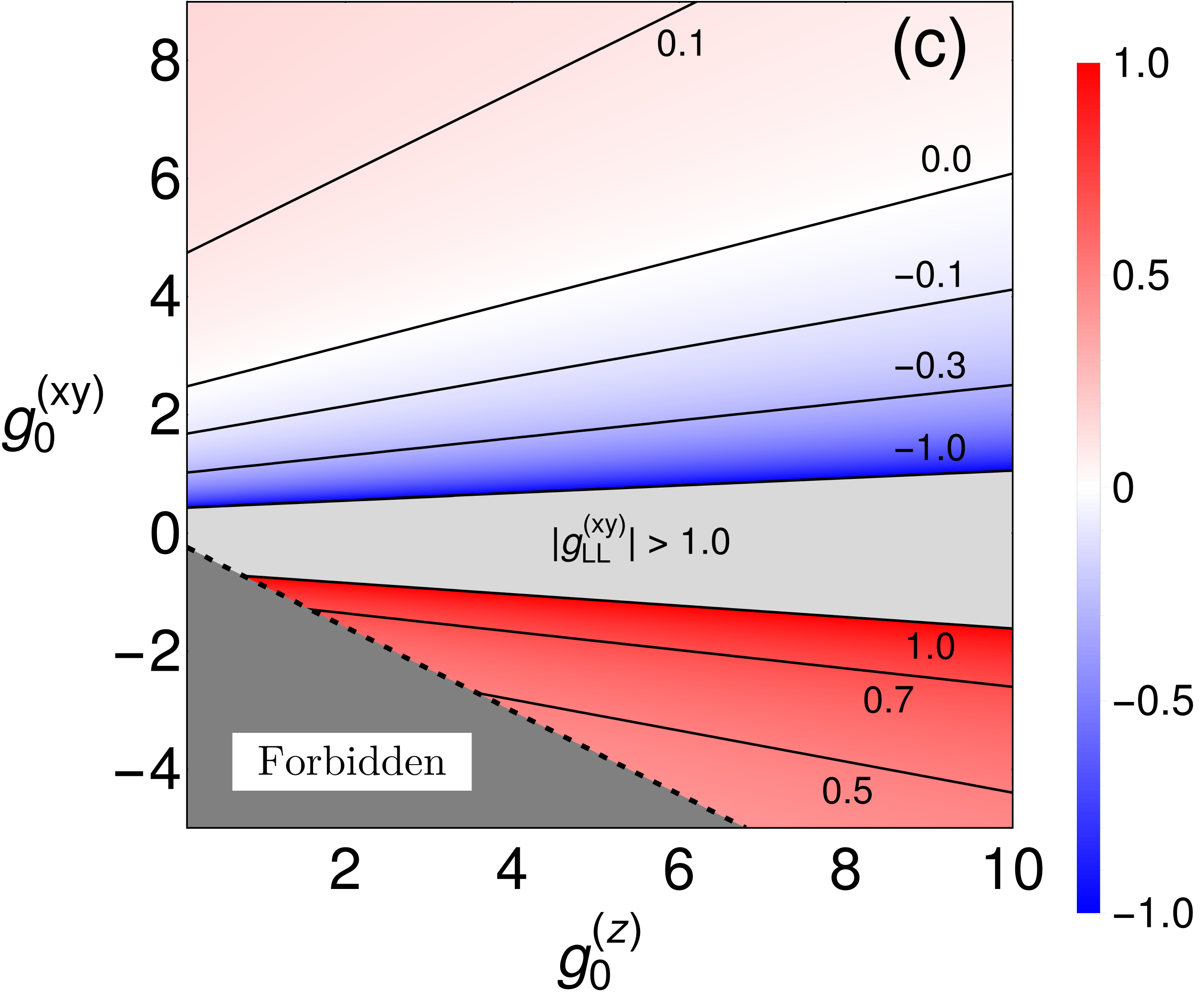}
\caption{Constraints on the interactions $v_{i} (q)$ [$i = z, xy$] dictated by transport measurements. The parameters $g^{(i)}_{0},\ g^{(i)}_{LL},\ \xi$ are defined in Eq.~\ref{eq:vi}. (a) $g^{(z)}_{LL}$ as a function of $g^{(z)}_{0}$ for different $\Bsa$ (the field at which the slope of $\Ds$ vs $\nu$ changes sign). 
Consistency with experiments rules out negative values of $g^{(z)}_{0}$. 
(b) $g^{(xy)}_{LL}$ versus $g^{(xy)}_{0}$ for different values of $g^{(z)}_{0}$. 
The cross marks the smallest value of $g^{(xy)}_{0}$ at which the theoretical model remains consistent with experiments. 
Notably, $v_{z}$ and $v_{xy}$ may comprise both repulsive and attractive components. 
(c) Contour plot of $g^{(xy)}_{LL}$ in the $g^{(z)}_{0}$--$g^{(xy)}_{0}$ plane. The dark gray region (below the dashed curve) is forbidden by experimental constraints. 
Here, we used $\xi = 0.3$ and $\e = 6$. In (b) and (c) we used $\Bsa = 26$ T and $\Bsb = 11$ T. }
\label{fig2}
\end{figure*}

A salient feature of the color map [inset of Fig.~\ref{fig1}(a)] is that at $B= 18$ T, $\Ds$ (defined as the average of $\Ds_{\pm}$) 
is monotonically decreasing with increasing $\nu$. This is not always the case:
Figure~\ref{fig1}(a) presents the variation of $\Ds$ with $B$ for different values of $\nu$. 
Strikingly, the various curves appear to cross around $B = \Bsa \sim 26$ T, implying that $\Ds(\nu)$ is an increasing (decreasing) function of $\nu$ for $B > \Bsa$ ($B < \Bsa$). 
Furthermore, $\Ds(\nu)$ decreases monotonically upon lowering $B$ and appears to vanish for sufficiently small $B$.
For example, $\Ds(\nu=2)$ vanishes at $B = \Bsb \sim 11$ T. This can be seen clearly in Fig.~\ref{fig1}(b), where the two resistance peaks observed for the high $B$ (which mark $\Ds_{\pm}$) merge into one at $B = \Bsb \sim 11$ T, implying $\Ds_{+} = \Ds_{-} = 0$. 

The theoretical challenge here is to account for the two most prominent features observed in the data: (a) the change in the slope of $\Ds$ vs $\nu$ from positive for $B > \Bsa$ to negative for $B < \Bsa$, and (b) the vanishing of $\Ds (\nu = 2)$ at $B = \Bsb$. 
A subsidiary puzzle is the nature of the $\nu = 2$ ground state as $D$ is tuned close to $\Ds$. 
Additionally, the theoretical model has to be consistent with previous observations at $\nu=0$, such as the canted antiferromagnet and layer polarized phases.

\textit{Theoretical Model.}~The LL spectrum of BLG close to charge neutrality (chemical potential $|\mu|\ll \hoc$) consists of eight nearly degenerate LLs, corresponding to the spin, valley and orbital degrees of freedom. Experimental evidence, e.g. the absence of any dependence on the in-plane field in the activation energy gaps measured at $\nu = 2, 3$~\cite{PKim_BLG_2010} and the relatively large effective Zeeman coupling~\cite{Jun_BLG_LLs}, indicate that in the filling factor range of interest to us ($\nu \sim 2$) the electronic states are
spin-polarized. We therefore restrict the Hilbert space in the model to four LLs, labelled by the orbital ($N = 0,1$) and valley ($\al = \pm$) indices. The two orbitals are not degenerate as there is no symmetry relating them. 
On the other hand, the two valleys are degenerate unless inversion symmetry is broken by a perpendicular electric field $D$ (or sublattice potentials, which are ignored here).  
The one-body part of the Hamiltonian is hence given by $H_{0} = \sum_{N \al k} \e_{\al N} \ROc_{N \al k} \LOc_{N \al k}$, where $k$ is the guiding center index in the Landau gauge, and
\begin{align} \label{eq:en}
  \e_{\al N} = N \D_{10} + \al \frac{\D_{D}}{2} |\MCP_{N \al}|. 
\end{align}
Here $\D_{10}$ is the energy gap between the two orbitals (for $D = 0$), $\MCP_{N \al}$ is the layer polarization, and $\D_{D} \propto D$ is the interlayer potential difference generated by $D$.  

To evaluate the energies and wave functions of the (non-interacting) states, we employed an effective four-band model (corresponding to the four sites of the unit-cell)~\cite{Falko_BLG_LLs}, which includes all tight-binding parameters found to be finite in ab-initio studies~\cite{JungMacDonald_BLG}.  
In particular, our model incorporates both trigonal warping and the hopping between  Bernal-stacked sites exactly (see Ref.~\onlinecite{SM} for details). 
Ignoring the Bernal-sites leads to perfect valley-layer locking, such that $\MCP_{N \al} = \al$. By contrast, in the full 4-component spinor the weight on these sites increases with $B$ and $\MCP_{N \al}$ depends on $N$~\cite{Hunt_BLG_2017}.  
This orbital dependence has significant impact on the variation of $\Ds$. Trigonal warping modifies the density profile of the wave functions at each site, which affects the interaction matrix elements and plays an important role in stabilizing novel ground states (see e.g. Ref.~\onlinecite{GM_BLG_2017}). 

The interacting part of the Hamiltonian comprises two components, $H_{c}$ and $H_{v}$. 
$H_{c}$ is an SU(4) symmetric (screened Coulomb) density-density interaction. 
The (Fourier transformed) pair potential for this is $v_{c} (q) = \frac{E_{c}}{\e} v_{\text{eff}}(q)$, where $E_{c} = \frac{e^{2}}{4\pi \e_{0}} \frac{1}{\ell}$ is the Coulomb energy scale, $\e$ is the relative permittivity of hBN, and $v_{\text{eff}}(q)=f(q)/q \ell$ where $f(q)$ is a form factor that accounts for screening from the top and bottom gates as well as higher energy LLs (at the RPA level)~\cite{SM}.  
$H_{v}$ represents the lattice-scale corrections, which reduce the valley symmetry to $U(1) \times \MBZ_{2}$. 
We assume that these corrections do not depend on the orbitals, 
and only include the terms present in the Kharitonov model of BLG~\cite{Kharitonov_BLG_2012,Kharitonov_BLG_PRB_2012} which may be expressed as 
\begin{align}
2 \pi \ell^{2} \times \frac{1}{2A} \sum_{i = x,y,z} \sum_{\Vq} v_{i} (q) : \rho_{i} (\Vq\,) \rho_{i}(-\Vq\,):,
\end{align}
where $A$ is the area of the sample, and 
$\rho_{i} (q)$ the (Fourier transform of) $i^{\text{th}}$ component of local isospin density~\cite{SM}. 
Valley $U(1)$ symmetry leads to $v_{x} (q) = v_{y} (q) \equiv v_{xy} (q)$. 
While $v_{i} (q)$ ($i = z, xy$) is replaced by a constant in the OM, here we assume the more general form, 
\begin{align} \label{eq:vi}
  v_{i}(q) = g^{(i)}_{0} \times \left( E_{c} \frac{a}{\ell} \right) \big[1 - g^{(i)}_{LL} \times \ka  \times e^{-\frac{\xi}{2} (q \ell)^{2}} \big] 
\end{align}
where $a$ is the lattice constant. In the limit $\ka \rightarrow 0$, $v_{i}(q)$ reduces to the standard short-range form with strength 
$g^{(i)}_{0} \left( E_{c} \frac{a}{\ell} \right) \propto B$. 
For finite $\ka$ the second term of (\ref{eq:vi}), which phenomenologically models corrections due to LL mixing, becomes progressively more important, with the characteristic scale $\ka \times \left( E_{c} \frac{a}{\ell} \right) \propto \sqrt{B}$. 
We emphasize that the two components of $v_{i}(q)$ differ not just in their range, but crucially also in their dependence on $B$, and may have different signs. 
The dimensionless numbers $g^{(i)}_{0}$, $g^{(i)}_{LL}$ and $\xi$, assumed to be independent of $B$, are the tuning parameters of the model.

\begin{figure}[t]
\centering
\includegraphics[width=\columnwidth]{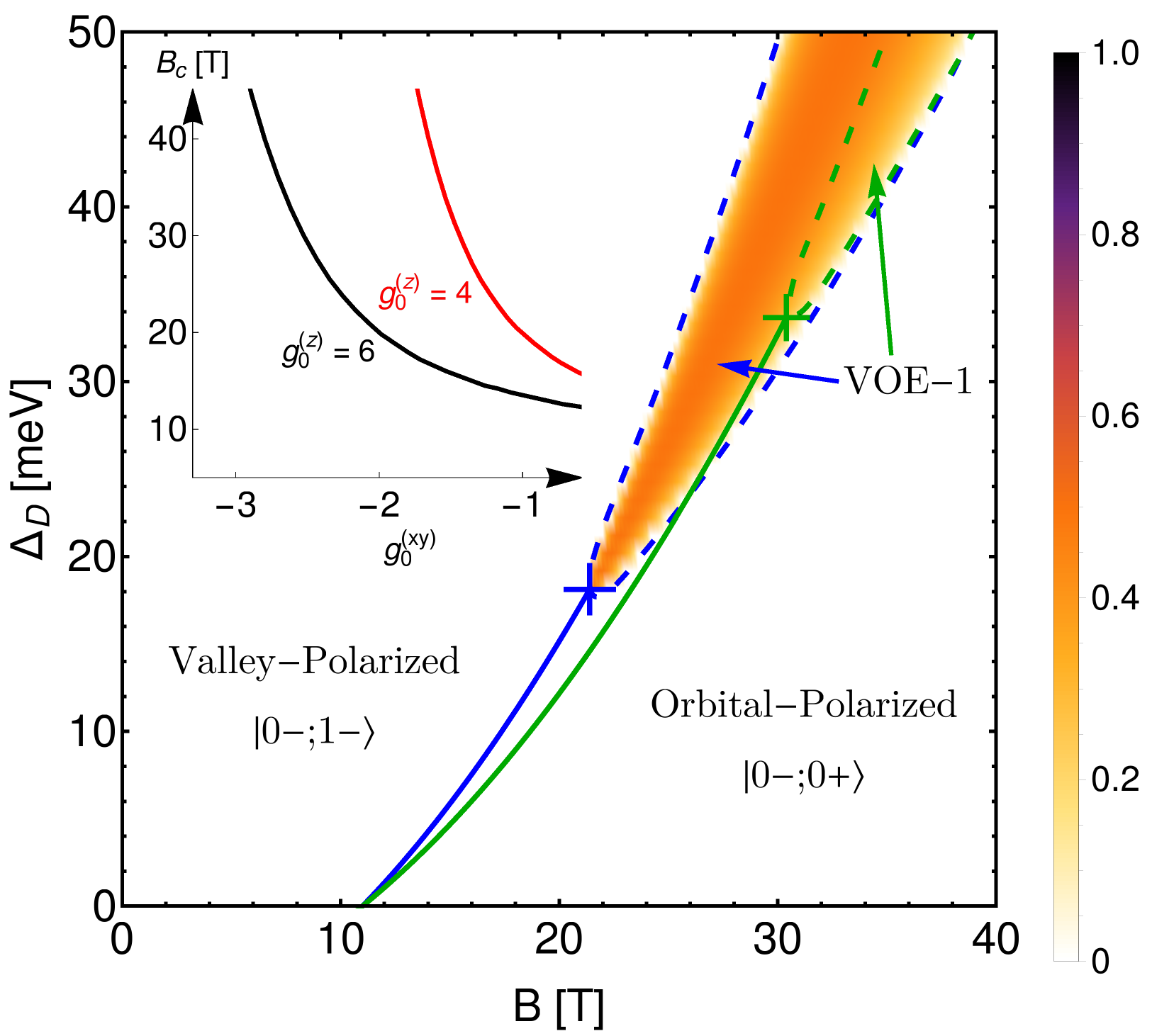}
\caption{Phase diagram at $\nu = 2$. 
The solid lines mark a first order transition between the valley-polarized and orbital-polarized phases (corresponding to $\Ds$);
the blue (green) curves correspond to $g^{(xy)}_{0} = -2.0$ ($g^{(xy)}_{0} = -2.5$). 
The plus signs (at $B = B_{c} \sim 20 T$ and $30 T$) mark a critical point where the first order phase boundary terminates. 
The dashed lines at higher $B$ mark the region where an intervening valley-orbital entangled phase (denoted by VOE-1) emerges around $\Ds$, which allows for a continuous transition between the two polarized phases. 
The color map shows the variation of the order parameter $\ANl \eta_{x} \ANr$ in the VOE-1 phase (see text) for $g^{(xy)}_{0} = -2.0$. 
The VOE-1 phase is characterized by a density matrix with the form in (\ref{eq:DeltaVOE}) with $\theta_{A} \in (0,\pi)$ and $\theta_{B} = 0$. 
The inset shows the variation of $B_{c}$ (the field at which the first order transition ends) with $g^{(xy)}_{0}$ for different $g^{(z)}_{0}$. 
Here, we used $\e = 6$, $\xi = 0.3$ and $g^{(z)}_{0} = 6.0$. 
$g^{(i)}_{LL}$ were chosen such that $\Bsa = 26$ T and $\Bsb = 11$ T.} 
\label{fig3}
\end{figure}

We treat the interactions in the self-consistent HF  approximation. 
The HF ground state, assumed to be translationally invariant, is characterized by the single-particle averages 
$\ANl \ROc_{N_{1} \al_{1} k_{1}} \LOc_{N_{2} \al_{2} k_{2}} \ANr = \delta_{k_{1} k_{2}} \D^{N_{1} \al_{1}}_{N_{2} \al_{2}}$, which minimize the variational energy. 
All details (including the $B$ dependence) of the interaction potentials and the single-particle wave functions are folded into the set of Hartree and Fock couplings~\cite{SM}.

\textit{Variation of $\Ds$.}~The $\nu = 2$ ground state corresponds to complete filling of two of the four LLs  included in the model. 
Equation~(\ref{eq:en}) suggests that the (non-interacting) ground state is  
$ |0-,0+ \ANr \equiv \Pi_{k} \ROc_{0-k}\ROc_{0+k} |0\ANr $ for $D \sim 0$ and a valley polarized phase $|0-,1- \ANr$ at large (and positive) $D$. 
The transition occurs at $D = \Ds(\nu=2)$ for which the energy of the two states is equal. 
Comparing the HF variational energy of these two states leads to an analytic equation for $\Ds(\nu=2)$~\cite{SM}. 

Upon reducing the filling factor to $\nu = 2 - \delta \nu$, the highest energy occupied LL is partially depleted. 
For $\delta \nu \ll 1$ this yields a linear equation $\Ds(2 - \delta \nu) = \Ds(2) - m_{\Ds} \delta \nu$, where the slope of $\Ds$ vs $\nu$ ($m_{\Ds}$) is 
\begin{align} \label{eq:mD}
  m_{\Ds} = \big(\MCF^{(c)}_{0000} - \MCF^{(c)}_{1111}\big) + \big(\MCF^{(z)}_{0000} - \MCF^{(z)}_{1111}\big).
\end{align}
Here $\MCF^{(i)}_{NNNN}$ is the Fock integral for Coulomb ($i = c$) and $v_{z}$ ($i = z$) interactions which couples electrons within one of the $|N, \al \ANr$ LL's~\cite{SM}. 
For repulsive interactions $\MCF^{}_{0000} \geq \MCF^{}_{1111}$ since the $N = 0$ states are more localized than those with $N = 1$. 
Hence, $m_{\Ds} > 0$ for all $B$ if only Coulomb interactions are present. 
In order to account for the experimental observations, $\MCF^{(z)}_{0000} - \MCF^{(z)}_{1111}$ must be sufficiently negative at $B < \Bsa$ and positive at higher $B$. 
We find that this cannot be achieved without a finite $g^{(z)}_{LL}$ [Fig.~\ref{fig2}(a)]. 
Our measurements constrain both $g^{(z)}_{0}$ and $g^{(z)}_{LL}$ to be positive, suggesting that $v_{z}$ must have both short-ranged repulsive and longer-ranged attractive components. 

Next, we turn to the vanishing of $\Ds(\nu=2)$ at $B = \Bsb \sim 11$ T. 
This can be achieved for generic values of $g^{(z)}_{0}$ and $g^{(xy)}_{0}$, if $g^{(xy)}_{LL}$ is also finite [Figs.~\ref{fig2}(b,c)]. 
Interestingly, the experimental results also constrain the possible values of the bare lattice interaction parameters ($g^{(z)}_{0}$ and $g^{(xy)}_{0}$). Specifically, $g^{(z)}_{0}$ may only assume positive values, while $g^{(xy)}_{0}$ must be larger than a certain cutoff [Fig.~\ref{fig2}(c)]. 

\textit{Intervalley Coherence.}~The analysis thus far considered ground states for which $\al$ and $N$ are good quantum numbers. 
Since two LLs with different valley and orbital indices are nearly degenerate in the $D \sim \Ds$ regime, the system may be able to lower the variational energy by hybridizing these LLs and forming a more complex ground state. 
We performed unrestricted HF calculations over a wide range of parameters to explore the nature of the $\nu = 2$ phase. 
This analysis uncovered a rich variety of possible ground states involving hybridization between different pairs of LLs~\cite{SM}. 
Here, we restrict the range of parameters to $g^{(z)}_{0} > 0$ and $g^{(xy)}_{0} < 0$, which is consistent with previous studies at $\nu = 0$ in this system~\cite{PKim_BLG_Nu0_2013,Jun_BLG_Nu0_2019,Jun_CAF_2021}. 
In this regime, the state is well-described for all $B$, $D$ by an ansatz for  $\D _{N_2 \alpha_2}^{N_1 \alpha_1}$ of the form 
\begin{align} \label{eq:DeltaVOE}
 \frac{1}{2} \left( \begin{array}{cccc} 
    1 + \cos(\theta_{A}) &   & 0 & \sin(\theta_{A}) \\
    0 & 1 + \cos(\theta_{B}) & \sin(\theta_{B}) & 0 \\
    0 & \sin(\theta_{B}) & 1 - \cos(\theta_{B}) & 0 \\
    \sin(\theta_{A}) & 0 & 0 & 1 - \cos(\theta_{A})
  \end{array} \right) 
\end{align}
in the ~$(N\alpha)=(0+, 0-, 1+, 1-)$ basis. The angles $\theta_{A,B} \in [0, \pi]$ parameterize the state~\cite{NPhi}. 
The orbital-(valley-)polarized state is described by $\theta_{A} = \theta_{B} = 0$ ($\theta_{A} = \pi$, $\theta_{B} = 0$). $\theta_{A,B} \neq 0, \pi$ correspond to  inter-orbital valley-coherent phases which smoothly interpolate between the two polarized states. These VOE phases break the $U(1)$-valley symmetry and are characterized by the order parameter $\ANl \eta_{x} \ANr = \frac{1}{2} \big[ \sin(\theta_{A}) + \sin(\theta_{B}) \big]$. 
Our analysis finds that for generic parameters (consistent with experiments), the polarized phases are separated by a first order transition for low values of $B$.  
The first order boundary terminates at a certain magnetic field, and a VOE phase appears in a finite parameter range around $D = \Ds$ for higher values of $B$ (Fig.~\ref{fig3} shows a typical phase diagram). We refer to this phase as VOE-1 because (generically) valley coherence emerges only in one of the sectors, i.e., $\theta_B = 0$ and $\theta_{A} \in (0,\pi)$ for $D > 0$. 
Such a valley coherent phase may exhibit Kekule BO.  
At the lattice scale these phases break translational symmetry, but upon coarse-graining to the scale $\ell$, they do not, in accordance with our original assumptions regarding $\D$. 

\textit{Discussion.}~The results presented above rely on the HF approximation, which ignores the effect of quantum fluctuations and correlations (beyond exchange). 
However, we believe that the qualitative features of our results would remain unaltered even when these effects are included. 
The cornerstone of our analysis is the distinct behavior of $\Ds (\nu)$ at low and high $B$-fields. 
Our measurements show that $\Ds (\nu)$ is a relatively smooth function for $1 < \nu \leq 2$. 
By contrast, correlation effects, which are crucial in stabilizing fractional QH phases, strongly depend on the precise value of $\nu$ and may be wildly different even for nearby fractions. 
This indicates that such effects do not play an important role in determining the \textit{qualitative} behavior of $\Ds (\nu)$ over a broad range of filling factors, which is apparently well-captured by the HF approximation. 
We emphasize that correlations beyond HF do affect $\Ds (\nu, B)$ quantitatively, even at higher $B$~\cite{Hunt_BLG_2017}. 

Our model further accounts for the vanishing of $\Ds(\nu=2)$ at $B = \Bsb$. 
Its experimental value ($\Bsb\sim 11\, {\rm T}$) allows us to constrain two of the four tuning parameters in the model, the couplings of the components arising from LL mixing ($g_{\text{LL}}$). 
The fact that even the qualitative behavior of the measured $\Ds$ cannot be explained without finite $g_{\text{LL}}$ strongly implies that LL-mixing plays a crucial role in determining the ground state, by introducing a effective attractive interactions that scale differently with $B$. These interactions become particularly pronounced at low $B$. 

We note that the U(1) valley symmetry is an artifact of the continuum approximation, and the restriction to just two-body interactions. LL mixing would not only modify the two-body potential, but also introduce three and higher-body terms. Since $3(\VK - \VK^{\prime})$ (where $\VK$ and $\VK^{\prime}$ are the locations of the valley centers in the Brillouin zone) is a reciprocal lattice vector, the lattice translation symmetry allows for three-body Umklapp terms transferring 3 fermions from one valley to the other. These terms reduce the U(1) symmetry, associated with the conservation of the difference of charge between the valleys, to $\MBZ_{3}$. Hence, the valley-coherent phase breaks a discrete symmetry, and may exist at finite temperatures. In fact, it corresponds to a Kekule bond-ordered phase, similar to those observed in STM experiments on MLG recently~\cite{STM_1,STM_2,STM_3}.    

\textit{Conclusions.}~Using high-quality BLG devices, we explored the behavior of the critical electric field $\Ds(\nu,B)$ 
in the range $1 < \nu \leq 2$, and observed a qualitative difference between the high and low $B$ regimes. 
Remarkably, we found that the standard theoretical models of BLG are not consistent with these measurements. 
Instead, it is crucial to consider the corrections to the lattice-scale interactions arising from LL-mixing, which we argued lead to an effective attraction at short but finite length scales.
We presented a phenomenological model of these which accounts for the experiments. 
It moreover predicts an inter-orbital valley-coherent phase for $D \sim \Ds$ at high $B$, 
which may be observed as a bond-ordered state in STM experiments. 
Our work motivates a detailed theoretical analysis of the LL-mixing corrections to lattice-scale interactions in MLG and BLG. 
Their effect on other integer and fractional QH states is another interesting direction for future investigations.  

\textit{Acknowledgements.}~We thank Chunli Huang, Ribhu Kaul and Benjamin Sacepe for useful discussions.
ES, HAF and GM thank the Aspen Center for Physics (NSF Grant No. 1066293) for its hospitality, and financial support by the US-Israel Binational Science Foundation through award No. 2016130. 
UK and ES acknowledge the support of the US-Israel Binational Science Foundation through award No. 2018726, and the Israel Science Foundation (ISF) Grant No. 993/19. 
HAF acknowledges the support of the NSF through Grant Nos. ECCS-1936406 and DMR-1914451.
KH and JZ acknowledge support from the National Science Foundation through Grant No.~NSF-DMR-1904986. 
The experiments were performed at the National High Magnetic Field Laboratory which was supported by the National Science Foundation through Grant No.~NSF-DMR-1644779 and the State of Florida. 
KW and TT acknowledge support from JSPS KAKENHI (Grant Nos.~19H05790, 20H00354 and 21H05233).

\onecolumngrid
\clearpage

\setcounter{affil}{0}
\setcounter{page}{1}
\renewcommand{\thefigure}{S\arabic{figure}}
\setcounter{figure}{0}
\renewcommand{\theequation}{S\arabic{equation}}
\setcounter{equation}{0}
\renewcommand\thesection{S\arabic{section}}
\setcounter{section}{0}

\title{Supplementary Material for ``Phase diagram of the $\nu = 2$ quantum Hall state in bilayer graphene''}
\author{Udit Khanna}
\affiliation{Department of Physics, Bar-Ilan University, Ramat Gan 52900, Israel}
\author{Ke Huang}
\affiliation{Department of Physics, The Pennsylvania State University, University Park, Pennsylvania 16802, USA}
\author{Ganpathy Murthy}
\affiliation{Department of Physics and Astronomy, University of Kentucky, Lexington, Kentucky 40506, USA}
\author{H.~A. Fertig}
\affiliation{Department of Physics, Indiana University, Bloomington, Indiana 47405, USA}
\author{Kenji Watanabe}
\affiliation{Research Center for Functional Materials, National Institute for Materials Science, 1-1 Namiki, Tsukuba 305-0044, Japan}
\author{Takashi Taniguchi}
\affiliation{International Center for Materials Nanoarchitectonics, National Institute for Materials Science, 1-1 Namiki, Tsukuba 305-0044, Japan}
\author{Jun Zhu}
\affiliation{Department of Physics, The Pennsylvania State University, University Park, Pennsylvania 16802, USA}
\author{Efrat Shimshoni}
\affiliation{Department of Physics, Bar-Ilan University, Ramat Gan 52900, Israel}

\begin{abstract}
This supplementary material provides additional details regarding our theoretical analysis. 
Section I describes the model in detail. 
The Hartree-Fock equations are derived in Section II, and Section III presents some additional results supplementing the ones provided in the main Letter. 
\end{abstract}

\maketitle

\begin{centering}
  \section{Theoretical Model}
\end{centering}

\begin{centering}
  \subsection*{Single-Particle States}
\end{centering}

At zero magnetic field, the low-energy electronic properties of bilayer graphene (BLG) may be 
accurately described by a 4-band continuum model~\cite{SMcCann_BLG,SJungMacDonald_BLG}. The 4 bands correspond to the 
4 sites $A, B, \PrA, \PrB$ in the unit-cell. Here $A, B$ ($\PrA, \PrB$) are the inequivalent sites in the 
bottom (top) layer, and $B, \PrA$ are the Bernal-stacked sites. 
In the basis $(A, B, \PrA, \PrB)$, the effective Hamiltonian (for each valley and spin index) is~\cite{SJungMacDonald_BLG}
\begin{align} \label{eq:sH0C}
  \left( \begin{array}{cccc}
    0 & \hbar v \pis & -\hbar v_{4} \pis & -\hbar v_{3} \pi \\
    \hbar v \pi & \Delta & t_{\perp} & -\hbar v_{4} \pis \\
    -\hbar v_{4} \pi & t_{\perp} & \Delta & \hbar v \pis \\
    -\hbar v_{3} \pis & -\hbar v_{4} \pi & \hbar v \pi & 0 
  \end{array} \right). 
\end{align}
Here, we defined the wavevector $\pi = \al q_{x} + i q_{y}$ ($\al = \pm 1$ labels the two valleys). 
The band structure is broadly governed by the two largest parameters $v$ and $\tpp$, namely the velocity of 
Dirac fermions in monolayer graphene (MLG) and the vertical hopping between Bernal-stacked sites, respectively. 
$v_{3}$ parameterizes the trigonal warping, which strongly affects the low energy band structure~\cite{SFalko_BLG_LLs}. 
$v_{4}$ and $\D$ are smaller parameters that break the particle-hole symmetry. 

A perpendicular magnetic field ($B$) is introduced in~(\ref{eq:sH0C}) through the Peierls substitution $\Vq \rightarrow 
\Vq + \frac{e}{\hbar} \VA$ (the electron charge is $-e$). 
We employ the Landau gauge, $\VA = -B y \Hx$, which leads to, 
$$\hbar \pi \rightarrow -\frac{\sqrt{2} \hbar}{\ell} \ROa_{k} \text{ for } \al = +1 \text{, } \hbar \pi \rightarrow \frac{\sqrt{2} \hbar}{\ell} \LOa_{k} \text{ for } \al = -1.$$ 
Here, $\ell = \sqrt{\hbar/eB} \approx 26 / \sqrt{B[T]}$ 
nm is the magnetic length, $\LOa_{k} = \frac{1}{\sqrt{2}} \big[ (y/\ell) - k  + i q_{y} \ell \big]$ is the 
lowering operator of a Gallilean harmonic oscillator centered around $y = q_{x} \ell^{2} = k \ell$, and $k$ labels the 
guiding center. These operators satisfy $[\LOa_{k}, \ROa_{k}] = 1$. 
Then the Hamiltonians $H_{\al}$ for the two valleys (for each spin and guiding center index) are given by,
\begin{align} \label{eq:sH0M}
  H_{-} &= \hoco \left( \begin{array}{cccc}
    0 & \ROa & -\lv \ROa & -\ltpp \LOa \\
    \LOa & \lDl \lp & \lp & -\lv \ROa \\
    -\lv \LOa & \lp & \lDl \lp & \ROa \\
    -\ltpp \ROa & -\lv \LOa & \LOa & 0 
  \end{array} \right), 
\end{align}
\begin{align}
  H_{+} &= \hoco \left( \begin{array}{cccc} \label{eq:sH0P}
    0 & -\LOa & \lv \LOa & \ltpp \ROa \\
    -\ROa & \lDl \lp & \lp & \lv \LOa \\
    \lv \ROa & \lp & \lDl \lp & -\LOa \\
    \ltpp \LOa & \lv \ROa & -\ROa & 0 
  \end{array} \right) . 
\end{align}
For brevity, we suppressed the guiding center index ($k$). 
The energy scale $\hoco = \sqrt{2} \frac{\hbar v}{\ell} \approx 31 \sqrt{B \text{[T]}}$ meV is the cyclotron gap in MLG. 
Substituting the values found in ab-initio calculations~\cite{SJungMacDonald_BLG}, the dimensionless parameters $\la_{i}$ entering $H_{\al}$ are, 
\begin{align}
  \lp &= \frac{\tpp}{\hoco} \approx \frac{12}{\sqrt{B \text{[T]}}} \xrightarrow{B = 10 \, \text{T}} 3.8 \\ 
  \ltpp &= \frac{v_{3}}{v} = 0.11 \,\, \forall \, B \\
  \lv &= \frac{v_{4}}{v} = 0.05 \,\, \forall \, B \\
  \lDl &= \frac{\D}{\tpp} = 0.04 \,\, \forall \, B. 
\end{align}
A perpendicular electric field ($D$) 
may be included in $H_{\al}$ through an additional term, 
\begin{align} \label{eq:sH0D}
  H_{D} &= \frac{\D_{D}}{2} \left( \begin{array}{cccc}
    -1 & 0 & 0 & 0 \\
    0 & -1 & 0 & 0 \\
    0 & 0 & 1 & 0 \\
    0 & 0 & 0 & 1 
  \end{array} \right). 
\end{align}
$\D_{D}$ is the interlayer bias induced by $D$, which was estimated to be $\D_{D} \text{ [meV] } \approx 0.13 \times D \text{ [mV/nm]}$ in Ref.~\cite{SJun_BLG_LLs}. 

The Landau spectrum, obtained by diagonalizing $H_{\al} + H_{D}$, has Landau levels (LLs) with energies $\sim \text{sign}(N) \hoc \sqrt{N (N-1)}$, 
where $N \in \MBZ$ labels the LLs and $\hoc = \hoco / \lp \approx 2.6 \, B \text{[T]}$ meV is the cyclotron gap in BLG. 
The set of eight LLs comprising the $N = 0, 1$ orbitals within each spin and valley sector forms the relevant low-energy subspace close to 
charge neutrality. We refer to this subspace as ZLL. We further restrict the Hilbert space to a 
single spin species, assuming that the quantum Hall (QH) phases close to $\nu = 2$ are fully spin-polarized~\cite{SPKim_BLG_2010}. 

Since $\lv, \lDl \ll 1$, and $\D_{D} \ll \hoc$ in the experimentally relevant regime of $B$ and $D$, 
we treat these terms perturbatively when evaluating the eigenfunctions and energies of the ZLLs. 
For $\lv = \lDl = D = 0$, the energy of all ZLLs is exactly zero. The corresponding eigenfunctions ($|\phi_{N \al} \ANr$) 
may be found analytically (for arbitrary $\ltpp, \lp$), yielding 
\begin{align} \label{eq:zmode1}
  |\phi_{N -} \ANr &= \frac{1}{\MCN_{N}} \left( \begin{array}{c} 
  |\Psi_{N -} \ANr \\ 0 \\ -(1/\lp) \LOa |\Psi_{N -} \ANr \\ 0 \end{array} \right),  \\ \label{eq:zmode2}
  |\phi_{N +} \ANr &= \frac{1}{\MCN_{N}} \left( \begin{array}{c} 
  0 \\ (1/\lp) \LOa |\Psi_{N +} \ANr \\ 0 \\ |\Psi_{N +} \ANr \end{array} \right). 
\end{align}
Here, $\MCN$ is the normalization constant, and 
\begin{align} \label{eq:zN}
  \MCN_{N} &= \sqrt{\ANl \Psi_{N \al} | \Psi_{N \al} \ANr + \frac{1}{\lp^{2}} \ANl \Psi_{N \al} | \ROa \LOa | \Psi_{N \al} \ANr}\;.
\end{align} 
For finite trigonal warping, $|\Psi_{N \al} \ANr$ is a superposition of the nonrelativistic harmonic oscillator states~\cite{SGM_BLG_2017}, 
\begin{align}
  |\Psi_{N \al k} \ANr &= \sum_{m = 0}^{\infty} \al^{m} A_{N m} |3 m + N, k \ANr, \,\, \text{ where} \label{eq:zmP} \\ 
  A_{N m} &= \frac{(3 \lt)^{m}}{\sqrt{(3m + N)!}} \frac{\G(m + 1/3 + N/3)}{\G(1/3 + N/3)}, \\ 
  \lt &= \ltpp \times \lp = \frac{1.3}{\sqrt{B[T]}} \xrightarrow{B = 10 \, \text{T}} 0.4. 
\end{align} 
The index $k$ was included in (\ref{eq:zmP}) for completeness. $|m, k \ANr$ refers to the $m^{\text{th}}$ eigenstate of the harmonic oscillator centered around $y = k \ell$. 
The coefficients in $|\phi_{N \al}\ANr$ are functions of $B$, due to the trigonal warping (controlled by $\lt \propto 1/\sqrt{B}$) 
and due to the finite weight on Bernal sites (controlled by $1/\lp$). Many previous studies have neglected (at least) one of these 
terms since the trigonal warping tends to dominate at low-magnetic fields~\cite{SGM_BLG_2017}, while the weight on Bernal sites 
is more relevant at higher $B$ ($\gtrsim 10$ T)~\cite{SHunt_BLG_2017}. Here, both are included as we are interested in a 
wide parameter window. For future convenience, we define normalized  
coefficients $u^{N \al}_{m \eta}$ (where $\eta$ labels the sublattice) as follows,
\begin{align} \label{eq:zA1}
  |\phi_{N - k} \ANr &= \sum_{m \geq 0} \left( \begin{array}{c} 
  u^{N-}_{m A} |m, k \ANr \\  0 \\ u^{N-}_{m \PrA} |m, k \ANr  \\ 0 
  \end{array} \right), \\  \label{eq:zA2}
  |\phi_{N + k} \ANr &= \sum_{m \geq 0} \left( \begin{array}{c} 
  0 \\ u^{N+}_{m B} |m, k \ANr \\  0 \\ u^{N+}_{m \PrB} |m, k \ANr  \\ 0 
  \end{array} \right). 
\end{align}
Note that the states $|\phi_{N \al}\ANr$ have finite weight on both layers. 
The $B$-dependent layer-polarization ($\MCP_{N \al}$) of these states is given by, 
\begin{align} \label{eq:zP}
  \MCP_{N \al} = \frac{\al}{\MCN^{\, 2}_{N}} \bigg[ \ANl \Psi_{N \al} | \Psi_{N \al} \ANr - 
  \frac{1}{\lp^{2}} \ANl \Psi_{N \al} | \ROa \LOa | \Psi_{N \al} \ANr \bigg]\; .
\end{align}
Since $\MCP_{N \al} \propto \al$, the interlayer bias $\D_{D}$ also acts as a valley Zeeman term. 
In the limit $\lp \gg 1$, the Bernal sites decouple from the low-energy sector, and $H_{\al}$ [in (\ref{eq:sH0M},\ref{eq:sH0P})] 
reduces to a $2 \times 2$ form. In this limit, (\ref{eq:zmode1},\ref{eq:zmode2}) reduce to the form derived in Ref.~\cite{SGM_BLG_2017}, and 
$\MCP_{N \al} \rightarrow \al$, i.e., independent of $N$. However, this limit is only justified when the magnetic field is very small 
($B \lesssim 1$ T). For the fields under consideration here, the finite weight on the Bernal sites induces a nontrivial $N$ 
 dependence in $\MCP$, which qualitatively affects our results concerning the behavior of $\Ds(\nu)$. 

\begin{figure}[t]
\centering
  \includegraphics[width=\columnwidth]{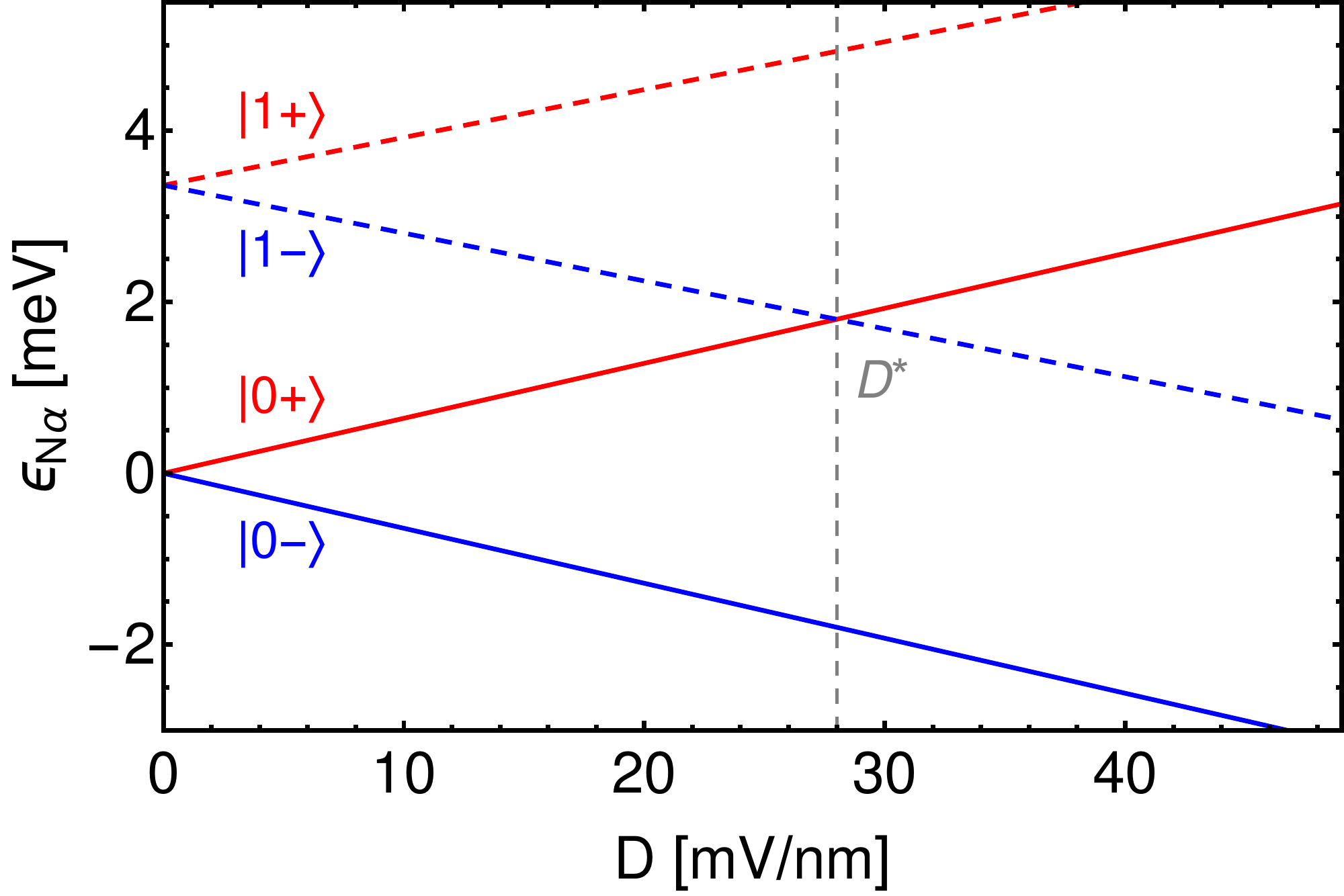}
  \caption{Variation of energy of the non-interacting LLs with the displacement field ($D$) at $B = 10$ T. 
  The solid (dashed) lines correspond to the $N = 0$ ($N = 1$) orbital, and red (blue) color corresponds to $\al = +1$ ($\al = -1$) 
  valley. The crossing of the $\e_{0 +}$ and $\e_{1 -}$ marks the critical field $\Ds$, at which a phase transition to a layer polarized 
  phase occurs.} 
\label{figS1}
\end{figure}

\begin{figure}[t]
\centering
  \includegraphics[width=\columnwidth]{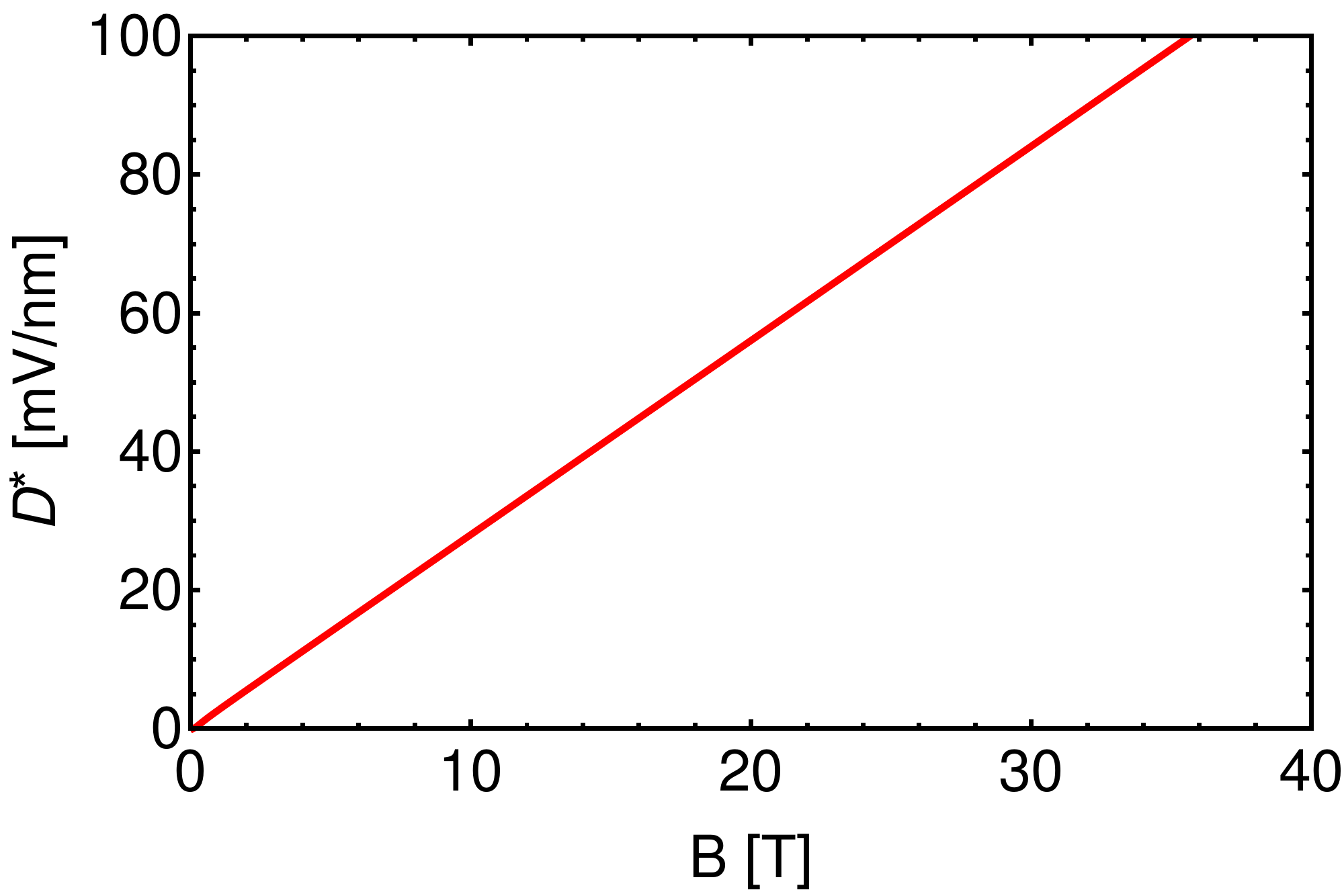}
  \caption{Variation of the critical field $\Ds$ with the magnetic field $B$ for non-interacting LLs. 
  Notably $\Ds$ is finite for all $B$. 
  Hence, the vanishing of $\Ds (\nu=2)$ at $B \sim 11 T$ observed in experiments must be an interaction effect. } 
\label{figS2}
\end{figure}

To first order in $\lv, \lDl, \frac{\D_{D}}{\hoc}$, the energy of these ZLL states is, 
\begin{align}
  \e_{N \al} = \frac{\D_{D}}{2} \MCP_{N \al} + \hoc \big( 2 \lv + \lDl \big) \MFN_{N}. 
\end{align}
$\MFN_{N} = \ANl \Psi_{N \al} | \ROa \LOa | \Psi_{N \al} \ANr / \MCN^{\, 2}_{N}$ is (roughly) the average Gallilean LL index of 
$| \Psi_{N \al} \ANr$. Adding an overall constant, the energies may be written in the form given in the main text, 
\begin{align} \label{eq:Zen}
  \e_{N \al} = \frac{\D_{D}}{2} \MCP_{N \al} + \D_{10} N, 
\end{align} 
where $\D_{10} = \hoc \big( 2 \lv + \lDl \big) \big( \MFN_{1} - \MFN_{0} \big)$. 
The modifications to the wave function due to finite $\lv, \lDl, \D_{D}$ are ignored in our analysis as these corrections 
(even at the first order) are very small in the relevant parameter regime, and do not affect the results qualitatively. 

Defining $\LOc_{N \al k}$ as the fermionic annihilation operator corresponding to the state $|\phi_{N \al k}\ANr$ as defined in 
(\ref{eq:zmode1},\ref{eq:zmode2}), the single-particle part of our effective Hamiltonian is, 
\begin{align}
  H_{0} &= \sum_{N = 0,1} \sum_{\al = \pm 1} \sum_{k} \e_{N \al}  \ROc_{N \al k} \LOc_{N \al k}, 
\end{align}
where $\e_{N \al}$ was given in (\ref{eq:Zen}) above. 
Notably the single-particle Hamiltonian $H_{0}$ does not contain any free parameters (other than $B$ and $D$). 
Figure~\ref{figS1} shows the evolution of these single-particle energies with $D$. 
The crossing point of $\e_{0 +}$ and $\e_{1 -}$ corresponds to $\Ds$, which varies linearly with $B$ (as shown in Fig.~\ref{figS2}). 
Crucially, the non-interacting $\Ds$ is finite at all $B$ and is independent of the filling factor $\nu$. 
Hence, the experimentally observed $\nu$ dependence of $\Ds$ and the vanishing of $\Ds(\nu=2)$ at $B \sim 11 T$ may only be explained through many-body effects. \\

\begin{centering}
  \subsection*{Interaction Hamiltonian}
\end{centering}

As described in the main text, the two-body interactions are divided in two parts: a long-range screened Coulomb interaction ($H_{c}$), and the lattice scale corrections ($H_{v}$). 
The interaction potential of the SU(4) symmetric $H_{c}$ has the form, 
\begin{align}
  v_{0} (\Vq \,) &= \frac{E_{c}}{\e} \ve (\Vq \,), \text{ where } \\ 
  E_{c} &= \frac{e^2}{4\pi \e_{0} \ell} \sim 55\sqrt{B\text{[T]}} \text{ meV}. 
\end{align}
Here, $\e$ is the relative permittivity of hBN, which we fix to be $6$ in our analysis~\cite{ShBN_DFT}.
In the absence of any screening, $\ve$ has the standard $1/q \ell$ form. In realistic samples, the Coulomb interaction is 
suppressed due to the metallic gates as well as the filled negative energy LLs. 
The metallic gates above and below the sample generate image charges in response to any fluctuation within BLG, which in turn 
modifies in the interaction potential. 
For simplicity, we assume both the gates are equidistant from the sample (at a distance of $d_{\text{gate}}$). 
Then the modified interaction is~\cite{SHunt_BLG_2017,SYoung_MLG_2021},
\begin{align}
  \tve (\Vq \,) &= \frac{1}{|\Vq \,| \ell} \tanh\big[|\Vq \,| d_{\text{gate}} \big]. 
\end{align}
Note that for $q d_{\text{gate}} \geq 1$, the gate-screening is not effective. As $B$ is reduced and $\ell$  
becomes comparable or larger than $d_{\text{gate}}$, the screening becomes much more pronounced (in the regime $q \ell \leq 1$), 
and effectively reduces $\tve$ to a short-range density-density interaction. 

We also include screening by the finite energy LLs at the RPA level. Then the combined effective interaction has the form~\cite{SGorbar_2010,SGorbar_2012}, 
\begin{align}
  \ve (\Vq \,) &= \frac{\tve (\Vq \,)}{1 + \tve (\Vq \,) \Pi (\Vq \,)}, \,\, \text{ where} \\
  \Pi (\Vq \,) &= a \times \tanh \big[ b \times ( |\Vq \,| \ell)^{2} \big]. \label{eq:sPI}
\end{align}
Here, $a$ and $b$ control the low and high $q$ behavior of $\Pi(q)$. The form in (\ref{eq:sPI}) fits very well to the 
(static) RPA polarization of the two-band model of BLG at $\nu = 0$~\cite{SPapic_BLG_2014}. The fitting procedure yields, 
$a_{\text{RPA}} \sim 5.54 \frac{E_{c}}{\e \hoc} $ and $b_{\text{RPA}} = 0.62$~\cite{SHunt_BLG_2017}. 
Employing the 4-band model and varying the filling factor away from 0 would likely lead to different values for $a, b$ and 
perhaps a different form for $\Pi(q)$~\cite{SKyrylo_BLG_2012}. 
We observed that our results do not depend very sensitively on the precise values of $a$ and $b$ (unless they deviate very significantly 
from the RPA values). In light of this, and in order to limit the number of free parameters, we fix $a = a_{\text{RPA}}$ and $b = b_{\text{RPA}}$. 

In principle, the interlayer and intralayer Coulomb interactions may have a different form. Here, we do not consider such effects 
to simplify our analysis. 
After projecting the screened Coulomb interactions onto the ZLL, we may write the Hamiltonian as, 
\begin{align}
  H_{c} &= 2 \pi \ell^{2} \times \frac{1}{2A} \sum_{\Vq} v_{0} (\Vq \,) : \rho(\Vq \,) \rho(-\Vq \,) : \, ,
\end{align}
where $\rho(\Vq\,)$ is the projection of (the Fourier transform of) the density operator onto the (spin-polarized) ZLLs, and $A$ is the sample area. 
Note that there are no free tuning parameters in $H_{c}$ within our analysis. 
The density operator may be written in terms of the fermion operators $\LOc_{N \al k}$ as, 
\begin{align}
  \rho(\Vq\,) = \sum_{\{N, \al \}, k} \ROc_{N_{1} \al_{1} k} &\LOc_{N_{2} \al_{2} k+q_{x}} 
  \times \nonumber \\
  &e^{-i q_{y} (k + \frac{q_{x}}{2}) \ell^{2}} \trho^{\al_{1} \al_{2}}_{N_{1} N_{2}} (\Vq \,), 
\end{align}
where the matrix element $\trho$ is, 
\begin{align} \label{eq:str0}
  \trho^{\al_{1} \al_{2}}_{N_{1} N_{2}} (\Vq\,)
  &= \sum_{\eta} \sum_{\{m\} \geq 0} \big( u^{N_{1} \al_{1}}_{m_{1} \eta} \big)^{*} u^{N_{2} \al_{2}}_{m_{2} \eta} \rho_{m_{1} m_{2}} (\Vq\,). 
\end{align}
Here, $\rho_{m_{1} m_{2}}$ is the corresponding matrix element in the basis of nonrelativistic LLs,
\begin{align}
  \rho_{m_{1} m_{2}} (\vq\,) = \sqrt{\frac{m_{<}!}{m_{>}!}} &e^{i \theta_{q} (m_{1} - m_{2})} e^{- \frac{q^{2} \ell^{2}}{4}} 
   \times \nonumber \\
  \left( -i \frac{q \ell}{\sqrt{2}} \right)^{|m_{1} - m_{2}|} &L_{m_{<}}^{|m_{1} - m_{2}|} \left[ \frac{q^{2} \ell^{2}}{2} \right] . 
\end{align}
Within the ZLL, the two valleys have support on different sublattices. Using this fact, it is easy to see that the matrix element
defined in (\ref{eq:str0}) is diagonal in the valley index. This is a consequence of using local lattice density operators in $H_{c}$. 

Next, consider the lattice scale corrections which reduce the valley symmetry from SU(2) to U(1)$\times \MBZ_{2}$. 
BLG allows for a large number of such interactions due to the orbital and sublattice degree of freedom. 
In order to keep the number of parameters under control, we assumed that these interactions do not have orbital dependence. 
To account for the sublattice, we define (local) isospin operators $\tau_{x,y,z}$, satisfying the Pauli algebra: $\tau_{i} \tau_{j} = \de_{ij} \MCI_{4} + i \e_{i j k} \tau_{k}$, as follows, 
\begin{align}
  \tau_{x} = \left( \begin{array}{cccc} 
    0 & 0 & 0 & 1 \\
    0 & 0 & 1 & 0 \\
    0 & 1 & 0 & 0 \\
    1 & 0 & 0 & 0 
  \end{array} \right)&, \, 
  \tau_{y} = \left( \begin{array}{cccc} 
    0 & 0 & 0 & i \\
    0 & 0 & -i & 0 \\
    0 & i & 0 & 0 \\
    -i & 0 & 0 & 0 
  \end{array} \right), \nonumber \\
  \tau_{z} = &\left( \begin{array}{cccc} 
    -1 & 0 & 0 & 0 \\
    0 & 1 & 0 & 0 \\
    0 & 0 & -1 & 0 \\
    0 & 0 & 0 & 1 
  \end{array} \right). 
\end{align}
Their action on the ZLL states is described by, 
\begin{align}
  \ANl \phi_{N_{1} \al_{1}} | \tau_{z} | \phi_{N_{2} \al_{2}} \ANr &= \de_{\al_{1}, \al_{2}} \de_{N_{1} N_{2}} \times \al, \\
  \ANl \phi_{N_{1} \al_{1}} | \tau_{x} | \phi_{N_{2} \al_{2}} \ANr &= \de_{\al_{1},-\al_{2}} \de_{N_{1} N_{2}} \times \MCH_{N} 
\end{align}
where
\begin{align} \label{eq:smch}
  \MCH_{N} = \frac{1}{\MCN_{N}^{2}} &\bigg[ \ANl \Psi_{N+} | \Psi_{N-} \ANr \nonumber \\ 
  &- \frac{1}{\lp^{2}} \ANl \Psi_{N-} | \ROa \LOa | \Psi_{N+} \ANr \bigg]   = \MCH_{N}^{*}. 
\end{align}
Clearly, $\tau_{z}$ labels the two valleys, while $\tau_{x}$ switches between them up to the orbital dependent constant $\MCH_{N}$.
The three ($i = x,y,z$) components of the (Fourier transform of) local isospin density may be defined as, 
\begin{align}
  \rho_{(i)}(\Vq\,) = &\sum_{\{N, \al \}, k} \ROc_{N_{1} \al_{1} k} \LOc_{N_{2} \al_{2} k+q_{x}}  
  \times \nonumber \\
  &e^{-i q_{y} (k + \frac{q_{x}}{2}) \ell^{2}} \times \trho^{(i)}_{N_{1} \al_{1} ; N_{2}\al_{2}} (\Vq \,). 
\end{align}
Here, $\trho^{(i)}$ is a generalization of (\ref{eq:str0}) which allows for off-diagonal terms in the valley-index. Specifically, 
\begin{align} \label{eq:str}
  \trho^{(i)}_{N_{1} \al_{1} ; N_{2}\al_{2}} (\Vq\,)
  = \sum_{\{\eta \}} &\sum_{\{m\} \geq 0} \rho_{m_{1} m_{2}} (\Vq\,) \times \nonumber \\
  &\big( u^{N_{1} \al_{1}}_{m_{1} \eta_{1}} \big)^{*} \big[ \tau_{i} \big]_{\eta_{1} \eta_{2}} 
  u^{N_{2} \al_{2}}_{m_{2} \eta_{2}}. 
\end{align}
Following previous works~\cite{SKharitonov_BLG_2012,SKharitonov_BLG_PRB_2012,SGM_BLG_2017}, we assume that the lattice scale 
interactions may be expressed as density-density terms in the three isospin channels, 
\begin{align}
  H_{v} = 2 \pi \ell^{2} \times \frac{1}{2 A} \sum_{i} \sum_{\Vq} v_{(i)} (\Vq\,) : \rho_{i}(\Vq\,) \rho_{i}(-\Vq\,) :, 
\end{align}
where $v_{(x)} = v_{(y)}$ due to the reduced symmetry. In the limit $\lp \gg 1$, the Bernal sites decouple from the problem 
and the valley index gets locked to the (remaining) sublattice. Then the reduced ($2 \times 2$) $\tau$ matrices, which act in the sublattice 
space, may equivalently be considered to act on the valley index directly. 

Typically, the lattice-scale terms are assumed to have a range of the order of the lattice constant, which leads to $v_{(i)} (\Vq\,)$ 
being essentially independent of $q$. As described in the main text, here we consider the lowest order corrections arising from LL mixing 
and use, 
\begin{align}
  v_{(i)}(q) = g^{(i)}_{0} \times \left( E_{c} \frac{a}{\ell} \right) 
  \big[1 - g^{(i)}_{LL} \times \ka  \times e^{-\frac{\xi}{2} (q \ell)^{2}} \big]. 
\end{align}
Here $\ka = E_{c} / \hoc$ controls the LL-mixing, and $E_{c} \frac{a}{\ell}$ sets the scale of these interactions.  
In principle, the interaction range ($\xi$) may depend on $i$ as well. We used the same $\xi$ for all interactions for simplicity. 
Consequently, $\xi$, $g^{(i)}_{0}$ and $g^{(i)}_{LL}$ (for $i = x, z$) are five free parameters in our analysis, whose 
values we shall try to infer using the experimental results. \\ 

\begin{figure*}[t]
\centering
  \includegraphics[width=0.29\textwidth]{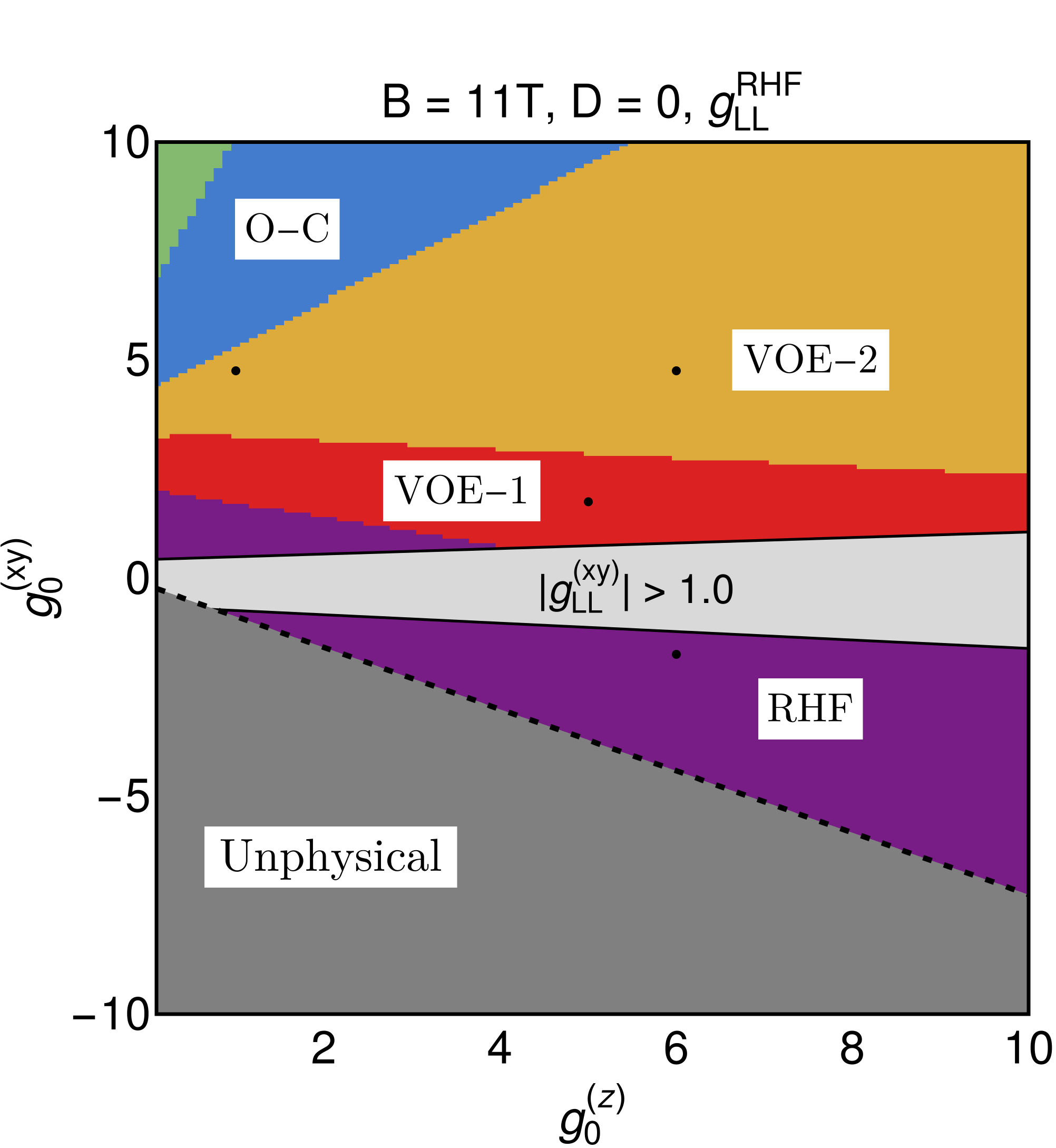}
  \includegraphics[width=0.29\textwidth]{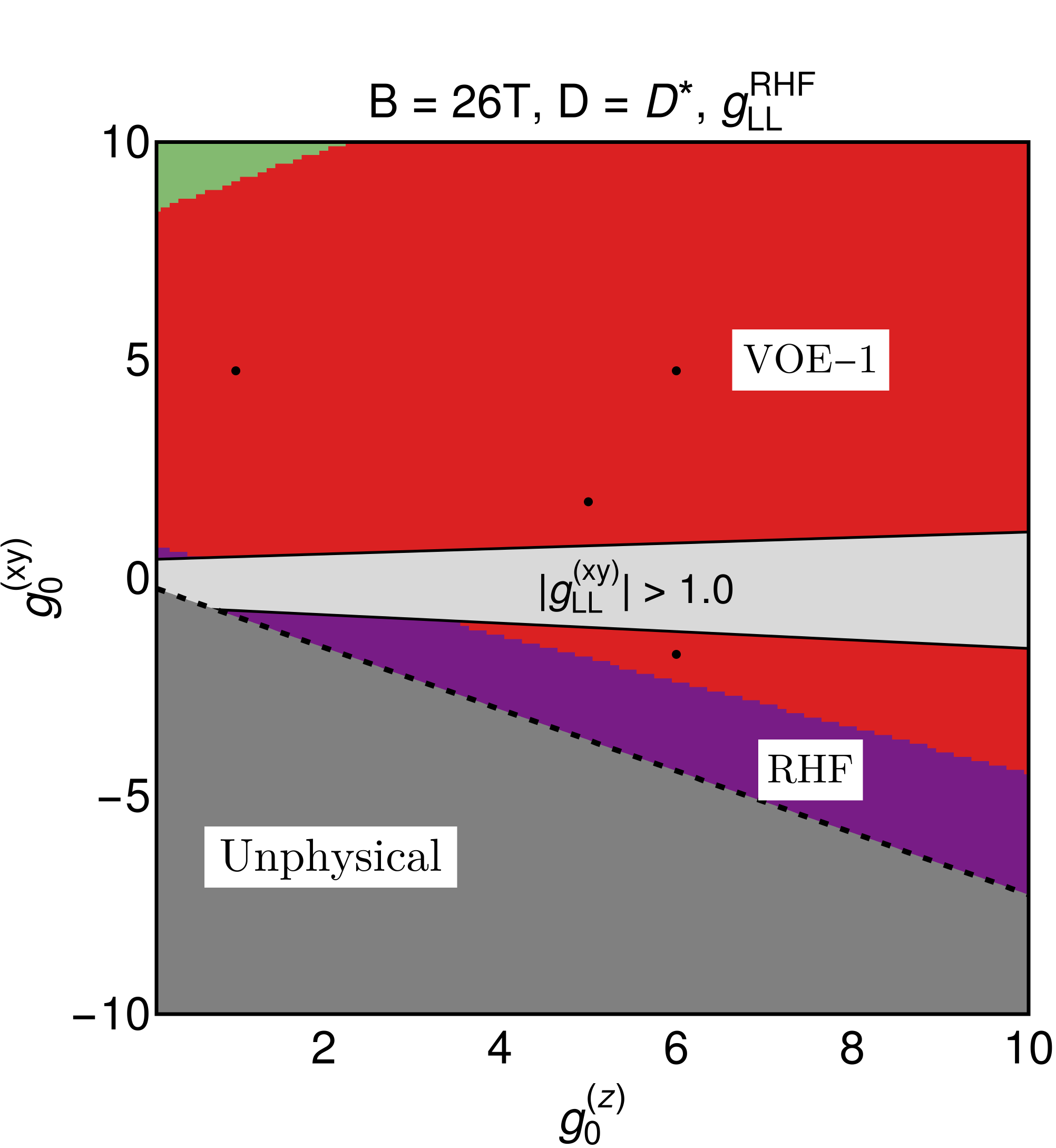}
  \includegraphics[width=0.32\textwidth]{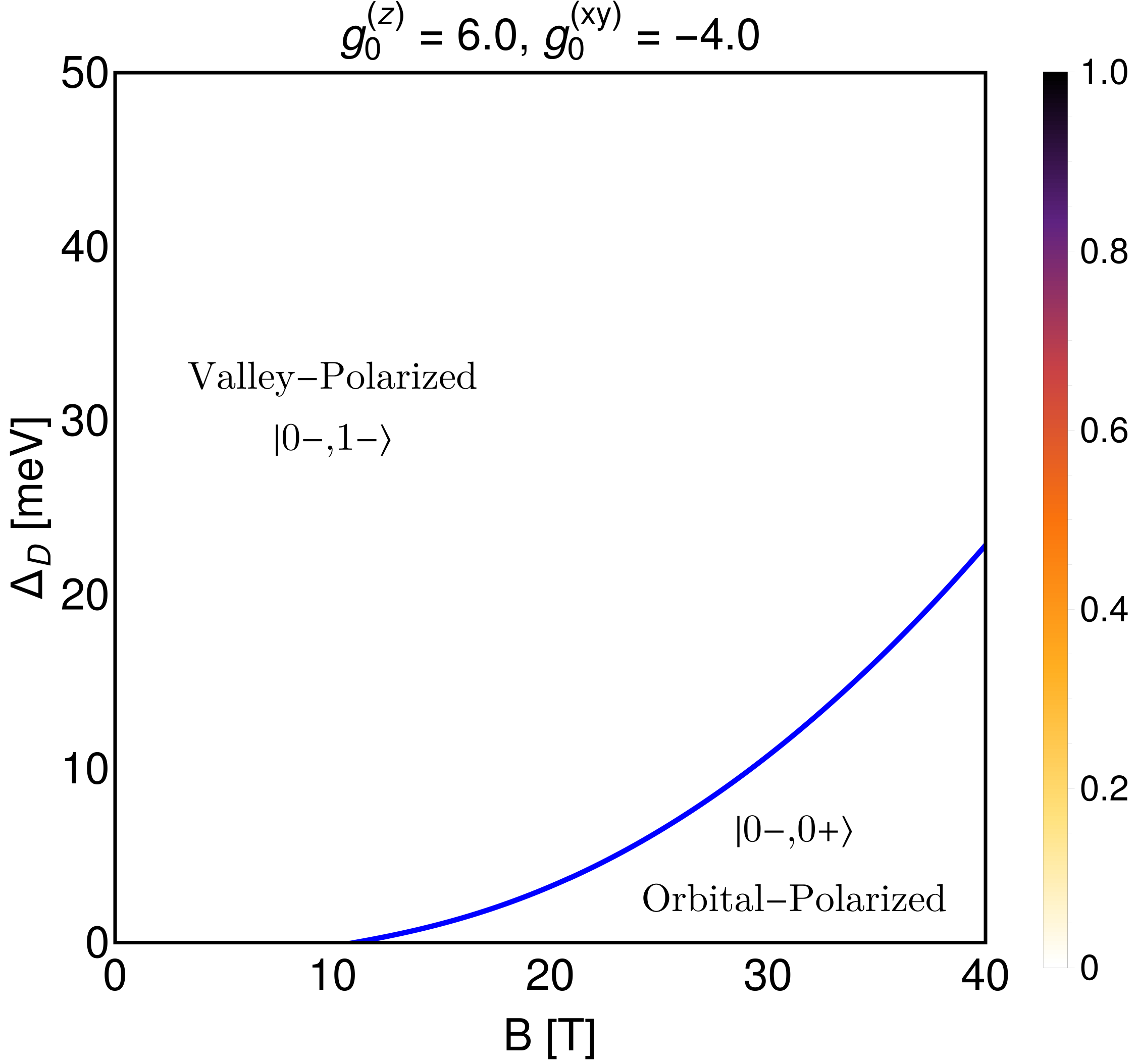}
  \includegraphics[width=0.31\textwidth]{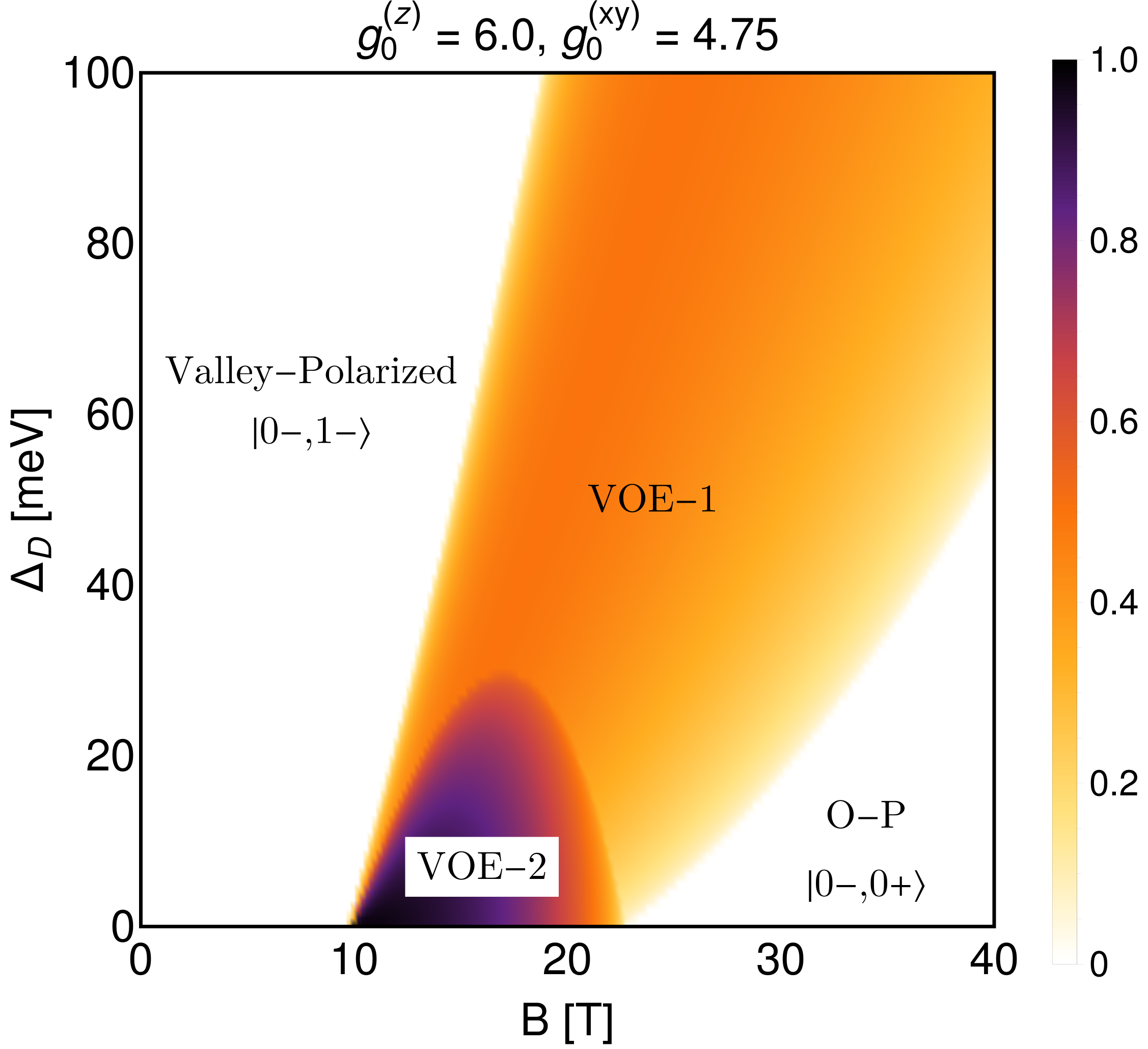}
  \includegraphics[width=0.31\textwidth]{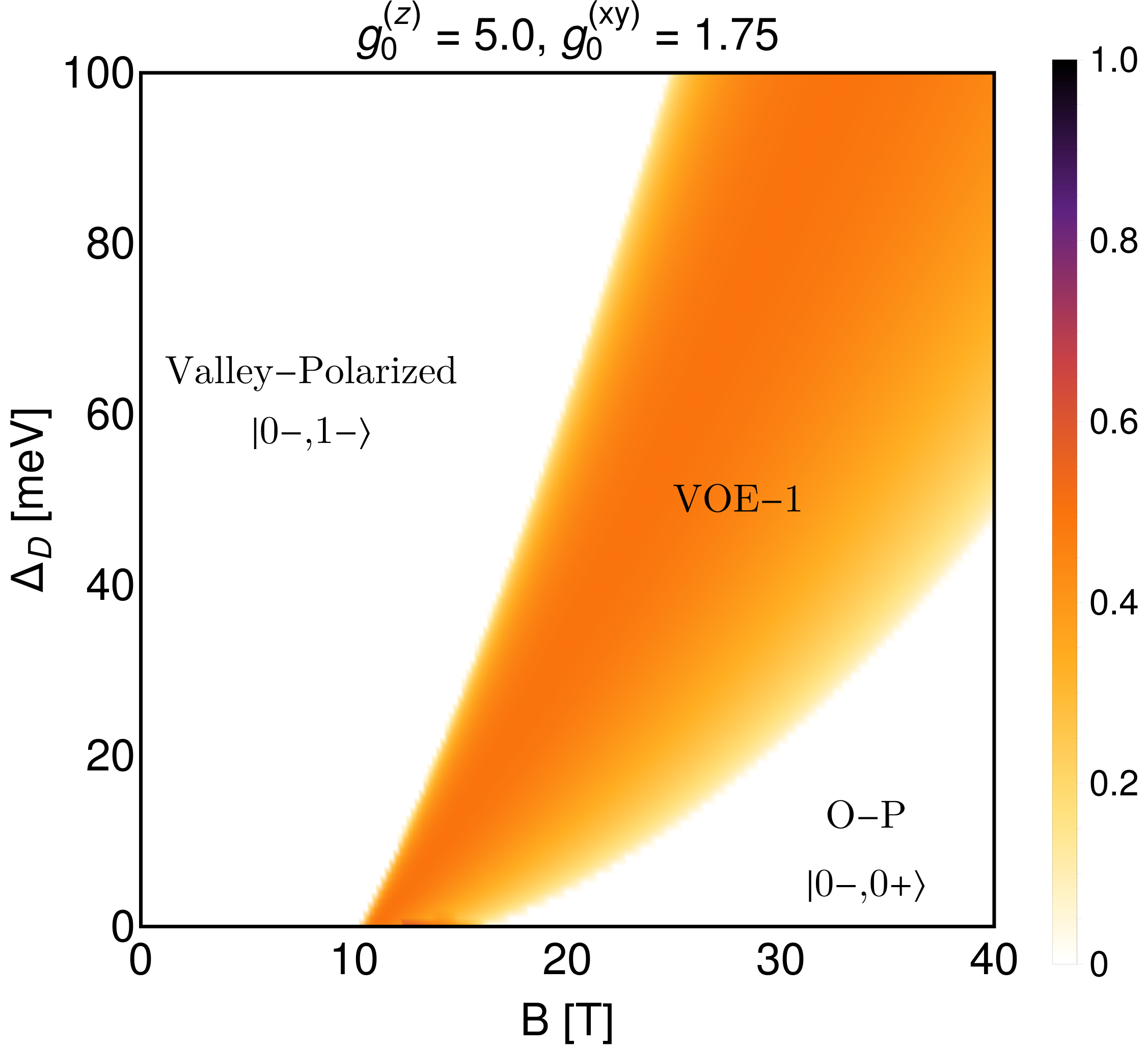}
  \includegraphics[width=0.34\textwidth]{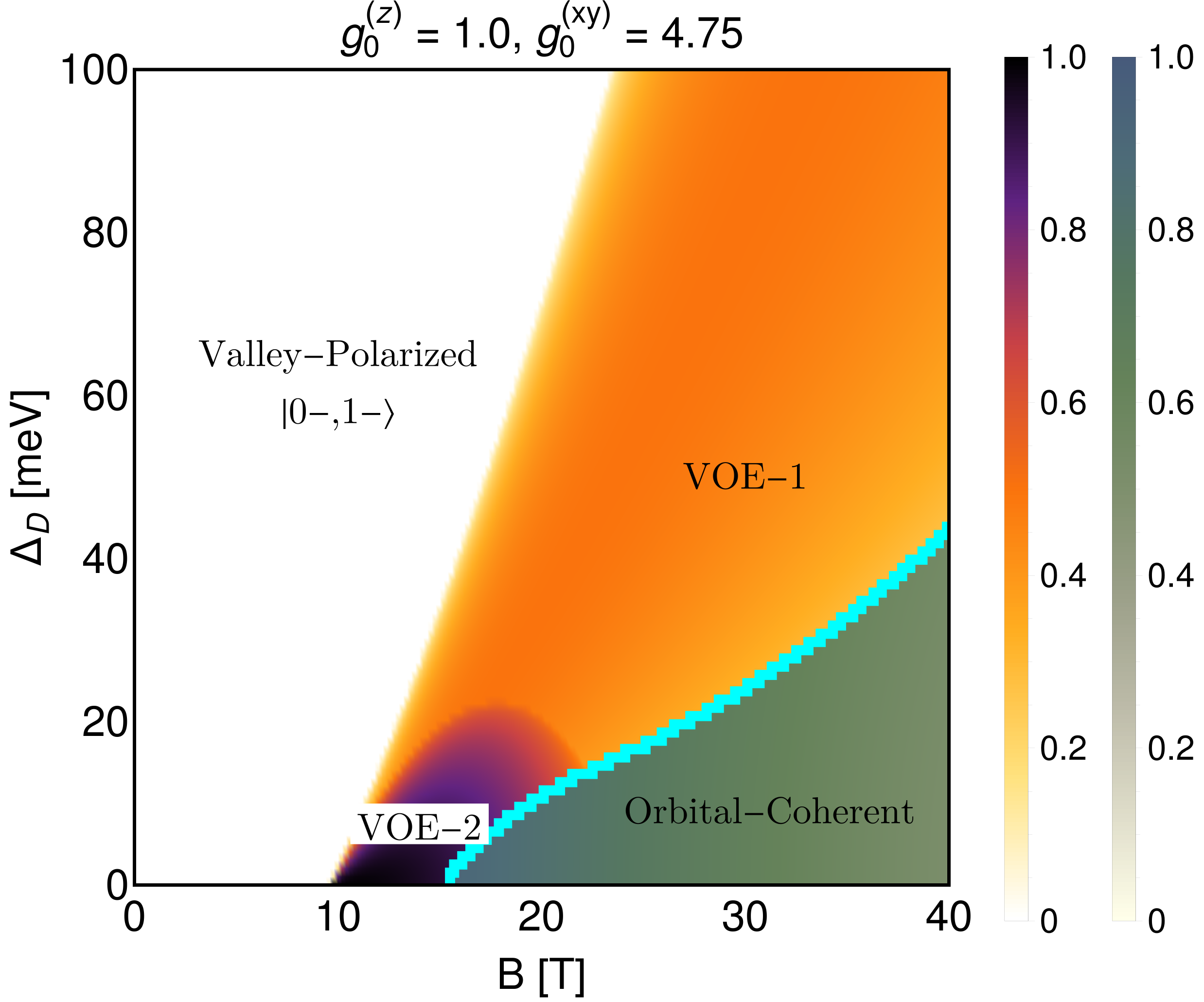}
  \caption{Complete phase diagram of the $\nu = 2$ QH state with modified interaction terms. 
  The first two panels of the top row depict the nature of the ground state at $B = 11$ and $26$ T and $D = \Ds$. 
  RHF (the purple region) marks the parameter space for which (orbital or valley) polarized phases are the lowest energy solutions. 
  VOE (red and yellow) marks the parameter space for which a valley-orbital entangled phase is the ground state. 
  VOE-1 (VOE-2) denotes the state in which one of (both) the angles $\theta_A$, $\theta_B$ (see Eq. (\ref{AB_anstaz})) parameterizing the density matrix is (are) non-trivial. 
  OC (blue region) corresponds to a (intra-valley) orbitally coherent phase. 
  The phase diagram in the $B$--$D$ plane depends on the values of the bare interaction parameters $g_{0}^{z,xy}$. 
  Figure 3 of the main text shows one of the possible phase diagrams for which $g_{0}^{xy} < 0$. 
  The remaining four panels show the other possibilities for different parameters. 
  In each case, the solid blue lines mark first order transitions and the color map shows the corresponding order parameter.} 
\label{figSP}
\end{figure*}

\vspace{5mm}

\begin{centering}
  \section{Hartree-Fock Analysis}
\end{centering}

We treat the interaction terms, $H_{c} + H_{v}$, in the self-consistent Hartree-Fock (HF) approximation. 
The HF ground state is characterized by the single-particle averages, 
$\ANl \ROc_{N_{1} \al_{1} k_{1}} \LOc_{N_{2} \al_{2} k_{2}} \ANr$, which are the variational parameters of the problem. 
We only consider states that do not break translation invariance along $x$ and $y$ (at the scale of $\ell$). Then only the averages that 
are diagonal in $k$ may be non-zero. Moreover, these averages are independent of $k$. The remaining variational parameters may be written as 
\begin{align}
  \ANl \ROc_{N_{1} \al_{1} k_{1}} \LOc_{N_{2} \al_{2} k_{2}} \ANr &= \delta_{k_{1} k_{2}} \D^{N_{1} \al_{1}}_{N_{2} \al_{2}} \nonumber \\
  &= \delta_{k_{1} k_{2}} \big( \D^{N_{2} \al_{2}}_{N_{1} \al_{1}} \big)^{*}. 
\end{align}\\

\begin{centering}
  \subsection*{HF Equations}
\end{centering}

In the HF approximation, $H_{c} + H_{v}$ is replaced by the corresponding Hartree and Fock terms. 
In a charge neutral system, the Hartree component of $H_{c}$ cancels out with the background positive charge. 
The Fock component has the form 
\begin{align}
  \sum_{\{N, \al\}, k} \Big[ V^{(c)}_{F; N_{1} \al_{1}, N_{2} \al_{2}} \ROc_{N_{1} \al_{1} k} \LOc_{N_{2} \al_{2} k} \Big]. 
\end{align}
Similarly, $H_{v}$ is replaced by, 
\begin{align}
  \sum_{N, \al, k} \Big[ &V_{H}^{(z)} \al \ROc_{N \al k} \LOc_{N \al k} + 
  V_{H; \al}^{(x)} \MCH_{N} \ROc_{N \al k} \LOc_{N \bar{\al} k} \Big] \nonumber \\
  + &\sum_{\{N, \al\}, k} \Big[ V^{(v)}_{F; N_{1} \al_{1}, N_{2} \al_{2}} \ROc_{N_{1} \al_{1} k} \LOc_{N_{2} \al_{2} k} \Big]. 
\end{align}
Here, $\MCH_{N}$ was defined in (\ref{eq:smch}), and $V_{H}$ and $V_{F}$ are the Hartree and Fock potentials. 
The Hartree potentials, related to the $q = 0$ component of the interaction potentials, are given by, 
\begin{align}
  V_{H}^{(z)} &= v_{(z)} (0) \times \sum_{N, \al} \al \D^{N \al}_{N \al}, \\
  V_{H; \al}^{(x)} &= 2 v_{(x)} (0) \times \sum_{N} \MCH_{N} \D^{N \bar{\al}}_{N \al}. 
\end{align}
For all interactions, the Fock potentials have the form, 
\begin{align}
  V^{(i)}_{F; N_{1} \al_{1}, N_{2} \al_{2}} &= - \sum_{N_{3}, N_{4}} \sum_{\al_{3}, \al_{4}} \MCF^{(i)}_{1, 4, 3, 2} \D^{N_{3} \al_{3}}_{N_{4} \al_{4}}. 
\end{align}
In the last equation, the index $i$ runs over the interactions ($c$, $z$ and $xy$), and the subscript $I = 1,2,3,4$ stands for $N_{I}, \al_{I}$. 
The Fock integral $\MCF^{(i)}$, which depends on the detailed form of the interaction potential, is given by, 
\begin{align}
  &\MCF^{(i)}_{1, 2, 3, 4} = \int \frac{d^{2} (q \ell)}{2 \pi} 
  v_{(i)} (q) \trho^{\al_{1} \al_{2}}_{N_{1} N_{2}} (q) \trho^{\al_{3} \al_{4}}_{N_{3} N_{4}} (-q). 
\end{align}
Using (\ref{eq:str}) in the definition above, it may be recast as 
\begin{align}
  \sum_{\{m \eta\}} &J^{(i)}_{m_{1} m_{2} m_{3} m_{4}} \Big[ 
  \big( u^{N_{1} \al_{1}}_{m_{1} \eta_{1}} \big)^{*} \big[ \tau_{i} \big]_{\eta_{1} \eta_{2}} 
  u^{N_{2} \al_{2}}_{m_{2} \eta_{2}} \Big] \times \nonumber \\
  &\Big[ \big( u^{N_{3} \al_{3}}_{m_{3} \eta_{3}} \big)^{*} \big[ \tau_{i} \big]_{\eta_{3} \eta_{4}} 
  u^{N_{4} \al_{4}}_{m_{4} \eta_{4}} \Big], \,  
\end{align}
where
\begin{align}
  &J^{(i)}_{m_{1} m_{2} m_{3} m_{4}} = \int \frac{d^{2} (q \ell)}{2 \pi} 
  v_{(i)} (q) \rho_{m_{1} m_{2}} (q) \rho_{m_{3} m_{4}} (-q).  
\end{align}
The assumed rotational invariance of the interactions simplifies the analysis by forcing many of the 256 $\MCF$ integrals (for each interaction) to vanish and introducing simple equalities among the rest. 
In the end, only eight unique Fock integrals need to be computed for the $i = c, z$ interactions: 
$\MCF^{(i)}_{0000}$, $\MCF^{(i)}_{1111}$, $\MCF^{(i)}_{0011}$, $\MCF^{(i)}_{0110}$, 
$\MCF^{(i)+-}_{0000}$, $\MCF^{(i)+-}_{1111}$, $\MCF^{(i)+-}_{0011}$ and $\MCF^{(i)+-}_{0110}$. 
Here, we used the convention (for $i = c, z$ only)
\begin{align}
  &\MCF^{(i)}_{N_{1}, N_{2}, N_{3}, N_{4}} = \int \frac{d^{2} (q \ell)}{2 \pi} 
  v_{(i)} (q) \trho^{+ +}_{N_{1} N_{2}} (q) \trho^{+ +}_{N_{3} N_{4}} (-q), \\
  &\MCF^{(i)+-}_{N_{1}, N_{2}, N_{3}, N_{4}} = \int \frac{d^{2} (q \ell)}{2 \pi} 
  v_{(i)} (q) \trho^{+ +}_{N_{1} N_{2}} (q) \trho^{- -}_{N_{3} N_{4}} (-q). 
\end{align}
For the $i = xy$ interaction, there are only four unique Fock integrals: 
$\MCF^{(xy)}_{0000}$, $\MCF^{(xy)}_{1111}$, $\MCF^{(xy)}_{0011}$, $\MCF^{(xy)}_{0110}$, defined as 
\begin{align}
  &\MCF^{(xy)}_{N_{1}, N_{2}, N_{3}, N_{4}} = \int \frac{d^{2} (q \ell)}{2 \pi} 
  v_{(xy)} (q) \trho^{+ -}_{N_{1} N_{2}} (q) \trho^{- +}_{N_{3} N_{4}} (-q). 
\end{align}

The variational energy of the HF states, defined as $\ANl H_{0} + H_{c} + H_{v} \ANr$, may be written in terms of the Hartree and Fock potentials defined above. Again, the rotational invariance of the interactions simplifies the form of this functional. 
The final functional may be decomposed into a sum of five terms, each of which is a function of different components of the single-particle density matrix, as shown below:
\begin{align}
  \MCE \big[ \{\D\} \big] = &E_{0} \big[ \D^{0+}_{0+}, \D^{0-}_{0-}, \D^{1+}_{1+}, \D^{1-}_{1-} \big] + \nonumber \\
  &E_{1} \big[ \D^{0-}_{0+}, \D^{0+}_{0-}, \D^{1-}_{1+}, \D^{1+}_{1-} \big] + \nonumber \\
 &E_{2} \big[ \D^{1+}_{0+}, \D^{1+}_{0+}, \D^{1-}_{0-}, \D^{0-}_{1-} \big] + \nonumber \\
  &E_{3} \big[ \D^{1-}_{0+}, \D^{0+}_{1-} \big] + 
  E_{4} \big[ \D^{1+}_{0-}, \D^{0-}_{1+} \big]\; .
\end{align}
$E_{0}$ is a function of just the occupations. $E_{3}$ ($E_{4}$) is a quadratic function of the modulus of $\D^{0+}_{1-}$ ($\D^{1+}_{0-}$). 
$E_{1}$ is a function of off-diagonal terms involving the same orbital and different valleys. 
$E_{2}$ is a function of coherences between different orbitals within the same valley.
These components are given by
\begin{widetext}
  \begin{align}
    E_{0} &= \sum_{N \al} \e_{N \al} \D^{N \al}_{N \al} + \frac{1}{2} v_{z}(0) 
    \Big[ \big( \D^{0+}_{0+} - \D^{0-}_{0-} \big)  +  \big( \D^{1+}_{1+} - \D^{1-}_{1-} \big)  \Big]^{2}
    -\frac{1}{2} \Big[ \big( \D^{0+}_{0+} \big)^{2}  +  \big( \D^{0-}_{0-} \big)^{2}  \Big] \Big[ \MCF^{(c)}_{0000} + \MCF^{(z)}_{0000} \Big] 
    \nonumber \\
   &\,\,\,\,\,\,\,\, -\frac{1}{2} \Big[ \big( \D^{1+}_{1+} \big)^{2}  +  \big( \D^{1-}_{1-} \big)^{2}  \Big] \Big[ \MCF^{(c)}_{1111} + \MCF^{(z)}_{1111} \Big]
    - \Big[ \D^{0+}_{0+} \D^{1+}_{1+} + \D^{0-}_{0-} \D^{1-}_{1-} \Big] \Big[ \MCF^{(c)}_{0110} + \MCF^{(z)}_{0110} \Big] 
    \nonumber \\
    &\,\,\,\,\,\,\,\, -2 \D^{0+}_{0+} \D^{0-}_{0-} \MCF^{(xy)}_{0000} -2 \D^{1+}_{1+} \D^{1-}_{1-} \MCF^{(xy)}_{1111} 
    -2 \Big[ \D^{0+}_{0+} \D^{1-}_{1-} + \D^{0-}_{0-} \D^{1+}_{1+} \Big] \MCF^{(xy)}_{0110},  \\
    E_{1} &= |\D^{0+}_{0-}|^{2} \Big[ 2v_{xy}(0) \MCH_{0}^{2} + \MCF^{(z)+-}_{0000} - \MCF^{(c)+-}_{0000} \Big] 
   + |\D^{1+}_{1-}|^{2} \Big[ 2v_{xy}(0) \MCH_{1}^{2} + \MCF^{(z)+-}_{1111} - \MCF^{(c)+-}_{1111} \Big] \nonumber \\
  &\,\,\,\,\,\,\,\,+2 \text{Re}\Big(\D^{1+}_{1-} \D^{0-}_{0+} \Big) \Big[ 2v_{xy}(0) \MCH_{0} \MCH_{1} + \MCF^{(z)+-}_{0110} - \MCF^{(c)+-}_{0110} \Big], \\
  E_{2} &= -\Big(|\D^{1+}_{0+}|^{2} + |\D^{1-}_{0-}|^{2} \Big) \Big[ \MCF^{(z)}_{0011} + \MCF^{(c)}_{0011}  \Big] 
  -4 \text{Re}\Big(\D^{1+}_{0+} \D^{0-}_{1-} \Big) \MCF^{(xy)}_{0011}, \\
    E_{3} &= |\D^{0+}_{1-}|^{2} \Big[ \MCF^{(z)+-}_{0011} - \MCF^{(c)+-}_{0011}  \Big], \,\, \text{ and } \\
  E_{4} &= |\D^{1+}_{0-}|^{2} \Big[ \MCF^{(z)+-}_{0011} - \MCF^{(c)+-}_{0011}  \Big]. 
  \end{align}\\
\end{widetext}

\begin{centering}
  \subsection*{Variation of $\Ds$}
\end{centering}

As noted earlier, the variation of $\Ds$ with $B$ and $\nu$ observed in experiments cannot be explained within a non-interacting picture. 
To account for interaction effects on $\Ds$, we evaluated the value of $\D_{D}$ for which the HF energy of the orbitally polarized state 
($|0-, 0+\ANr$) is equal to the energy of the valley polarized state ($|0-, 1- \ANr$). 
This corresponds to the $\Ds (\nu = 2)$ within a restricted HF (RHF) analysis. 
A straightforward calculation gives, 
\begin{align}
  \Ds(\nu = 2) &= m_{\Ds} + 2 \D_{10} + 4 v_{z} (0) + 4 \MCF^{(xy)}_{0000} \nonumber \\ 
  &\,\,\,\,\,\,\,\,- 2 \Big( \MCF^{(c)}_{0110} + \MCF^{(z)}_{0110} \Big), \\
  m_{\Ds} = \Big( &\MCF^{(c)}_{0000} - \MCF^{(c)}_{1111} \Big) + \Big( \MCF^{(z)}_{0000} - \MCF^{(z)}_{1111} \Big). 
\end{align}
Here, we have absorbed an unimportant factor in the definition of $\Ds$. 
Next, in order to find the variation of $\Ds$ as the filling factor is reduced from $2$, we repeat the same calculation for 
variational states in which $\D^{0-}_{0-} = 1$ and the occupation of $|1-\ANr$ or $|0+ \ANr$ is $1 - \delta \nu$. 
These correspond to the zero temperature limit of thermal states in which the highest occupied LL is partially, but uniformly, occupied. 
Clearly, correlation effects (beyond HF) would stabilize fractional QH states at specific fillings. 
As explained in the main text, we expect that the gross qualitative variation of $\Ds$ with $\nu$ would not be controlled by such corrections. 
Within these approximations, we find, 
\begin{align}
  \Ds(\nu = 2 - \delta \nu) = \Ds(\nu = 2) -  m_{\Ds} \delta \nu. 
\end{align}
This implies that the slope of $\Ds$ vs $\nu$ is given by $m_{\Ds}$. As explained in the main text, 
this allows us to fix some of the interaction parameters using the experimental observations. 

\vspace{5mm}

\begin{centering}
  \section{Complete Phase diagram at $\nu = 2$}
\end{centering}

To find the complete phase diagram of the $\nu = 2$ state, we performed fully unrestricted HF calculations over a wide parameter window. 
Our results suggest that the $\nu = 2$ state can be described at any parameter, by one of the three different 2-angle ansatzes for $\D _{N_2 \alpha_2}^{N_1 \alpha_1}$ given below. Interestingly, these represent the most general density matrices in the $E_{0} + E_{1}$, $E_{0} + E_{2}$ and $E_{0} + E_{3} + E_{4}$ subsectors of the complete energy functional $\MCE$ respectively. In the ~$(N\alpha)=(0+, 0-, 1+, 1-)$ basis, these are 
\begin{align}
 \frac{1}{2} \left( \begin{array}{cccc} 
    1 + \cos(\theta_{0}) &  \sin(\theta_{0}) & 0 & 0 \\
    \sin(\theta_{0}) & 1 - \cos(\theta_{0}) & 0 & 0 \\
    0 & 0 & 1 + \cos(\theta_{1}) & e^{-i \phi} \sin(\theta_{1}) \\
    0 & 0 & e^{i \phi} \sin(\theta_{1}) & 1 - \cos(\theta_{1})
  \end{array} \right) ,
\end{align}
\begin{align}
 \frac{1}{2} \left( \begin{array}{cccc} 
    1 + \cos(\theta_{+}) &  0 & \sin(\theta_{+}) & 0 \\
    0 & 1 + \cos(\theta_{-}) & 0 & e^{-i \phi} \sin(\theta_{-}) \\
    \sin(\theta_{+}) & 0 & 1 - \cos(\theta_{+}) & 0 \\
    0 & e^{i \phi} \sin(\theta_{1}) & 0 & 1 - \cos(\theta_{-})
  \end{array} \right) ,
\end{align}
\begin{align}
\label{AB_anstaz}
 \frac{1}{2} \left( \begin{array}{cccc} 
    1 + \cos(\theta_{A}) &  0 & 0 & \sin(\theta_{A}) \\
    0 & 1 + \cos(\theta_{B}) & \sin(\theta_{B}) & 0 \\
    0 & \sin(\theta_{B}) & 1 - \cos(\theta_{B}) & 0 \\
    \sin(\theta_{A}) & 0 & 0 & 1 - \cos(\theta_{A})
  \end{array} \right) \,.
\end{align}
We find the lowest energy solution for each of these ansatzes by direct minimization of the corresponding energy functional with respect to the free parameters (i.e., the angular variables appearing in each ansatz.) 
The ground state is the lowest energy solution among these three. 
This ground state may be characterized by the order parameters
\begin{align}
  \ANl \tau_{x} \ANr &= \frac{1}{2} \Big[ \sin(\theta_{0}) + \sin(\theta_{1}) \Big]\; , \\
    \ANl \sigma_{x} \ANr &= \frac{1}{2} \Big[ \sin(\theta_{+}) + \sin(\theta_{-}) \Big]\; , \\
      \ANl \eta_{x} \ANr &= \frac{1}{2} \Big[ \sin(\theta_{A}) + \sin(\theta_{B}) \Big] \; .
\end{align}
Figure~\ref{figSP} shows the complete phase diagram at $\nu = 2$.


\begin{thebibliography}{47}
\bibitem [{NPh()}]{NPhi}%
  \BibitemOpen
  \href@noop {} {}\bibinfo {note} {In principle, the ansatz allows for an
  additional parameter corresponding to the relative phase between $A$ and $B$
  sectors. Here, we ignore this phase as it drops out of the energy
  functional~\cite{SM}.}\BibitemShut {Stop}%
\end{thebibliography}
\end{document}